\documentclass[useAMS,usenatbib]{mn2e}

\usepackage{graphicx}
\usepackage{times}
\usepackage{natbib}
\usepackage{color}
\usepackage{todonotes}
\newcommand{\cosmodeep}{\it{\small COSMODEEP}}

\newcommand{\apjs}{ApJS}

\newcommand{\apjl}{ApJL}

\newcommand{\aap}{A \& A}

\newcommand{\enzo}{\it{\small ENZO}}

\begin{document}
 
\title[DDA for radio astronomy]{Convolutional Deep Denoising Autoencoders for Radio Astronomical Images}
\author[C. Gheller, F. Vazza]{C. Gheller$^1$, F. Vazza$^{3,2,1}$\thanks{E-mail: claudio.gheller@inaf.it}\\
$^{1}$Istituto di Radio Astronomia, INAF, Via Gobetti 101, 40121 Bologna, Italy\\
$^{2}$ Hamburger Sternwarte, Gojenbergsweg 112, 21029 Hamburg, Germany\\
$^{3}$ Dipartimento di Fisica e Astronomia, Universit\'{a} di Bologna, Via Gobetti 92/3, 40121, Bologna, Italy
}

\date{Received / Accepted}
\maketitle
\begin{abstract}
We apply a Machine Learning technique known as Convolutional Denoising Autoencoder to denoise synthetic images of state-of-the-art radio telescopes, with the goal of detecting the faint, diffused radio sources predicted to characterise the radio cosmic web. In our application, denoising is intended to address both the reduction of random instrumental noise and the minimisation of additional spurious artefacts like the sidelobes, resulting from the aperture synthesis technique. The effectiveness and the accuracy of the method are analysed for different kinds of corrupted input images, together with its computational performance. Specific attention has been devoted to create realistic mock observations for the training, exploiting the outcomes of cosmological numerical simulations, to generate images corresponding to LOFAR HBA 8 hours observations at 150 MHz. Our autoencoder can effectively denoise complex images identifying and extracting faint objects at the limits of the instrumental sensitivity. The method can efficiently scale on large datasets, exploiting high performance computing solutions, in a fully automated way (i.e. no human supervision is required after training). 
It can accurately perform image segmentation, identifying low brightness outskirts of diffused sources, proving to be a viable solution for detecting challenging extended objects hidden in noisy radio observations. 

\end{abstract}

\label{firstpage} 

\section{Introduction}
\label{sec:intro}

The incoming generation of astrophysical observatories at various wavelengths is anticipated to have dramatically improved performance in terms of sensitivity and resolution, allowing the finding and characterisation of novel classes of sources, so far barely detectable. 
The diffused matter component of the universe represents a primary target for astrophysics. This is mainly composed by the Warm-Hot Intergalactic Medium (WHIM, \citealt{1999ApJ...514....1C}), already identified in the clusters of galaxies outskirts \citep[e.g.][]{2015Natur.528..105E} or between closely interacting cluster pairs (\citealt{2021A&A...647A...2R}), and predicted by numerical simulations as major building block of sheets and filaments, tracing the cosmic web (\citealt{2001ApJ...552..473D}). The high energy tail of the WHIM, made of ultrarelativistic charged particles (cosmic rays) is expected to emit synchrotron radiation, by spinning in the cosmic magnetic fields. The resulting emission is extremely faint due to the low particle density (fractions of the cosmological critical density), and the weak magnetic fields (whose magnitude is still uncertain, but $\ll 1\mu$G \citealt[e.g.][]{vern20}). Brightest sources are expected to have flux densities not exceeding $10^{-4}$ Jy at the limit of the sensitivity of current radio interferometers, like LOFAR \citep{2013A&A...556A...2V}, MeerKAT\footnote{https://www.sarao.ac.za/gallery/meerkat/}, MWA\footnote{https://www.mwatelescope.org/}, ASKAP\footnote{https://www.atnf.csiro.au/projects/askap/index.html}, which can only start scraping the surface of the cosmic web in the radio band \citep[e.g.][]{vern17,go19,vern20,bott20,bo21}. A substantial leap in the characterisation of such objects is expected within the next decade with the advent of the SKA (MID and LOW)\footnote{https://www.skatelescope.org/}. 
Instruments like LOFAR and MWA have also led to the identification of remnant radio galaxies \citep[e.g.][]{2017MNRAS.471..891G}, an additional  class of faint, extended radio sources, which often can reach angular and physical scales comparable to those of extended radio emissions connected to the WHIM. These sources, that are characterised by a steep radio emission spectrum and results from long-lasting evolutionary processes in radio galaxies, are detected at the limits of the sensitivity of the adopted radio telescope \citep[e.g.][]{ma20,hod21,ge21,du21} . 

The common feature of these observations is that they all target extended sources that are hard to detect due to their weak emission. A key requirement for the identification and the characterisation of such sources is the availability of effective methodologies able to distil their faint signal from the noise and the disturbances that affect the data and that can have different origin and features, depending on the context. 


In the case of radio interferometry, the aperture synthesis approach is exploited to generate observations in the radio band. Aperture synthesis consists in the adoption of sophisticated data processing techniques in order to obtain sky images from the collected signals, estimating the energy flux density coming from a given region of the sky, at a specific frequency. Besides random noise (of, e.g., thermal or electronic origin), these observations are affected by the resolution of the instrument, by the geometric pattern characterising the observation process (sparse antennas rotating with the Earth) and by a number of other effects related to the propagation of radio signals through the Earth's atmosphere (e.g. ionospheric disturbances or Radio-Frequency Interference) which may contaminate the final product. For a detailed overview of all these aspects, we refer to \citet{1999ASPC..180.....T}. 

Errors have to be properly treated in order to minimise their impact on the data. Any possible ``correcting'' process, however, can itself affect the enclosed information. Therefore, tailored strategies have to be evaluated in order to understand both their effectiveness and their final influence on the physical and statistical properties of the output data. In radio interferometry, data correction is generally performed in various steps (often with an iterative procedure). Firstly, the calibration step aims at estimating and correcting for time, frequency, and direction dependent instrumental errors (\citealt{2011A&A...527A.106S}, \citealt{2011A&A...527A.107S}). Various software tools address calibration, like DPPP (\citealt{2018ascl.soft04003V}) or KillMS\footnote{https://github.com/saopicc/killMS)}.
This is followed by imaging, that is the processes of Fourier-transforming the calibrated visibilities into images (for example DDFacet, \citealt{2018A&A...611A..87T} and WSClean, \citealt{offringa-wsclean-2014}). Then deconvolution (see \citealt{cornwell:1999ASPC..180.....T}) corrects the resulting images for the incomplete sampling of the Fourier plane. Deconvolution is implemented in terms of the Clean algorithm (\citealt{1974AAS...15..417H}) in its different variants (\citealt{1980AA....89..377C}, \citealt{1984AJ.....89.1076S}, among the most widely used), or other approaches, like the Maximum Entropy Model (\citealt{1985A&A...143...77C}), the MORESANE model (\citealt{2015AA...576A...7D}), the Purify algorithm (\citealt{purify}), the SASIR method (\citealt{2015JInst..10C8013G}), the RESOLVE Bayesian method (\citealt{2016AA...586A..76J}) and the clean multiscale deconvolution approach (\citealt{2011A&A...532A..71R}, \citealt{offringa-wsclean-2017}). Finally, denoising and source detection and characterisation are performed. The former can be accomplished adopting various techniques, like gaussian, FFT based or wavelet filtering. A comprehensive review with a performance comparison of different solutions is given by \citet[][]{2020A&A...643A..43R}. Denoising tools are usually available within source finding software packages, like PyBDSF (\citealt{2021A&A...645A..89M}), SoFiA (\citealt{2015MNRAS.448.1922S}) and AEGEAN (\citealt{2012MNRAS.422.1812H}) among the most up-to-date. Also of note are comprehensive and widely adopted software platforms, able to perform most of the previous tasks, like CASA (\citealt{2007ASPC..376..127M}), AIPS (\citealt{Greisen2003}), MIRIAD (\citealt{1995ASPC...77..433S}) and ASKAPsoft (\citealt{2020ASPC..527..591W}). 

The scenario outlined above is already challenging enough. However it will  be made even more complicated by
the flurry of data that future instruments, like the SKA, will soon produce. The full SKA is expected to deliver between 100 and 500 PBytes per year of science ready data.
Already its precursors and pathfinders (the aforementioned LOFAR, MeerKAT, MWA and ASKAP radio telescopes) are  generating datasets whose processing requires hardware and software resources at the limits of current technological capabilities. Therefore, next generations of software solutions will have to handle such big quantities of data. This has two main consequences: $i$) the software will have to be able to effectively exploit High Performance Computing (HPC) resources, in terms of parallelism, heterogeneous architectures, hierarchical memory and storage devices, advanced software solutions; $ii$) at the same time it will have to be automated, so to minimise human intervention in data processing, that would be impossible for such big data volumes and/or would introduce dramatic bottlenecks in the processing workflow. 

In this work we have investigated the adoption of Artificial Intelligence based methodologies with the goal of addressing the challenge of removing noise and artefacts from images resulting from the imaging and deconvolution processes, exploiting at the same time HPC solutions on data mimicking the LOFAR observations, in view of extending the approach to the upcoming results of the SKA. Our key objective is to develop an accurate and effective tool to minimise noise and other sources of contamination in real radio data, supporting precise image segmentation to fully identify extended radio sources even at levels of instrumental noise, capable of running on data of ``any'' size, scaling on  increasingly bigger and more capable computing architectures in a fully automated way (i.e. with no need of human supervision).

More specifically, we have explored Deep Learning solutions, that have proved to be effective for tasks relating to signal processing, computer vision, text analysis, speech recognition (\citealt{726791}, \citealt{NIPS2012_4824}, \citealt{DBLP:journals/corr/SimonyanZ14a}, \citealt{43022}, \citealt{DBLP:conf/cvpr/HeZRS16}, \citealt{garcia17}), among others. Deep Learning is a branch of Machine Learning that has become increasingly popular in the last decade, thanks to two concurrent factors: the availability of enough computing power to cope with complex, multi-layered neural networks, and the availability of enough data to perform the training. In the last years, it has also been adopted for a large number of applications in astronomy and cosmology. The most common usage is for classification tasks (see \citealt{2021arXiv210601571C}, \citealt{2021arXiv210600187K}, \citealt{2021arXiv210412980F} for the most recent examples), but also generative networks start to be widely adopted (see \citealt{2021arXiv210604014C}, \citealt{2021arXiv210512149B}, \citealt{2021PNAS..11822038L}, again for the latest examples). Deep Learning has also been exploited for the detection of sources, as in \citet{2021arXiv210607660S} or \citet{2021arXiv210600187K} and as in our previous work \citep[][]{2018MNRAS.480.3749G}, where we have explored the potential of Convolutional Neural Networks (CNN) in identifying the faint radio signal from extended cosmological radio sources (such as emission from shocked gas around galaxy clusters and filaments) in noisy radio observations, which are at the limit of sensitivity of instruments like ASKAP or LOFAR. The resulting methodology, named {\cosmodeep}, allowed us to detect diffuse radio sources and to localise their position within large images thanks to a tiling based procedure with an accuracy of around the 90\% (i.e. around 10\% of the detections are misclassified). Unsurprisingly, one of the outcomes of our study is that the accuracy of the CNN approach tends to drop as the signal to noise ratio decreases to values below 1 (although less than with other standard approaches), hence an effective preliminary image denoising would be decisive for improving the quality of the results. 

Various examples of image denoising methods based on Deep Learning techniques are available. For an exhaustive up-to-date review we refer to \citet{Tian2020DeepLO}. In particular, CNN based algorithms, like convolutional Deep Denoising Autoencoders (DDA) have proved to be useful for complex noisy images, audio or video streams (see, e.g., \citealt{2005ARA&A..43..139P}). The autoencoder methods have also been successfully applied to data segmentation problems (see e.g., \citealt{Ronneberger2015}, \citealt{6909475}). We have adopted such approach for our radio interferometry data. The detailed description of the adopted autoencoder architecture is given in Section \ref{sec:dda}.


The most important requirement for the usage of a Machine/Deep Learning based approach is the availability of data for the training of the network. In our case, this implies that a sufficiently large number of training images, representative of the target data, has to be available. The training set accounts for the ``corrupted'' images, that are the input to the DDA, together with the corresponding ``ground true'' counterparts (same images but with no errors), to compare the input images with. The method is unsupervised, so no labelling is needed. This is a huge advantage compared to supervised Deep Learning approaches, since labelling is, in general, a demanding, time consuming and error prone procedure. On the other hand, in radio astronomy, sufficiently large training sets of observational data are not available. Mock radio observations have to be generated starting from numerical simulations. Reproducing actual observations with high fidelity is however an outstanding task. In principle all the necessary physics should be included in the simulation. For instance, in the radio band, at low surface brightness the sky contribution is expected to be likely produced by both the synchrotron emission from the shocked cosmic web \citep{va15survey} as well as from the contribution from radio galaxies naturally forming in different environments \citep[][]{2021arXiv210607901H}. Only by including both, it is possible to produce a self-consistent simulation of the evolution of radio galaxies interacting with their environment, including their large-scale impact on the distribution of magnetic fields and on relativistic particles in filaments and voids. Our procedure to produce comprehensive models is described in the first part of Section \ref{sec:data}. 

In addition, these models have to be ``observed''. This means that as many random and systematic perturbations affecting the ideal image as possible should be included. This has been addressed as described in the second part of Section \ref{sec:data}. Mock Sky models and observed images make up an open access archive of the simulated radio images, which we make fully available (see Section \ref{data}).
In Section \ref{sec:results}, we present an extensive analysis of the performance of our DDA applied to different families of data, addressing the aforementioned challenges of data denoising and segmentation. The results are integrated in Appendices A and B by the specific study of how architectural and hyperparameters (input parameters of the network) choices affects the effectiveness of the autoencoder. Section \ref{sec:performance} summarises instead the computational performance of the DDA. 

The outcomes and main achievements of our study are discussed in Section \ref{sec:discussion}, followed by a conclusive summary, presented in Section \ref{sec:conclusions}.





\section{The Convolutional Deep Denoising Autoencoder}
\label{sec:dda}

An {\it autoencoder} is a type of neural network which aims at learning an identity map for the input data through encoding and decoding data pairs:
\begin{equation}
    \tilde I = D(E(I)),
\end{equation}
where $I$ and $\tilde I$ are the input and reconstructed images respectively, and $E$ and $D$ are the encoder and decoder transforms, which minimise the difference of the reconstructed and the input data:
\begin{equation}
    {D,E = \rm argmin_{(D,E)}} ({\rm MSE}(I-D(E(I))),
\end{equation}
where MSE is the mean squared error, defined as:
\begin{equation}
    {\rm MSE} (X-Y) = {1\over N} \sum_{i=1}^N (X_i - Y_i)^2,
\label{eq:mse}
\end{equation}
$X$ and $Y$ being two $N$-components vectors. 

The minimisation is accomplished by error back-propagation, optimising the model parameters of the encoder and the decoder through an iterative procedure called {\it training}.

In the case of a {\it denoising} autoencoder the minimisation is performed by comparing the reconstructed image calculated using as input a noisy or corrupted version $I_c$ of $I$, to the true image itself, i.e.:
\begin{equation}
    {D,E = \rm argmin_{(D,E)}} ({\rm MSE}(I-D(E(I_c))).
\end{equation}

\begin{figure*}
\begin{center}
\includegraphics[width=0.99\textwidth]{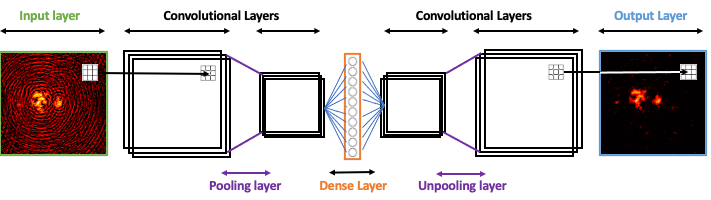}
\caption{Example of Deep Denoising Autoencoder architecture made of the input and output layers, two convolutional plus one pooling layers for downsampling, a specular set-up for upsampling and a dense central layer (the {\it latent space}).}
\label{fig:architecture}
\end{center}
\end{figure*}

A {\it convolutional deep} denoising autoencoder is a particular type of denoising autoencoder whose architecture is exemplified in Figure \ref{fig:architecture}. The encoder consists of an input layer, a downsampling convolutional network (composed by convolutional and pooling layers) and of a low-dimensional fully connected layer, (called {\it latent space}). The decoder starts with the fully connected layer, followed by an upsampling convolutional network, specular to the downsampling one, and an output layer returning images of the same size of the input ones.

During the downsampling sweep, the input data is processed by linear convolution:
\begin{equation}
    s_{m,n}^f = \sum_{i,j=-k}^{k} w_{i,j}^f p_{m+i,n+j} + b^f,
\end{equation}
where $p_{m,n}$ is the $(m,n)$ pixel of the input image, $w_{i,j}^f$ and $b^f$ are the weights and the bias of the convolutional kernel (also known as receptive window), whose size has been set in our model to 3$\times$3 pixels. Finally, $s_{m,n}^f$ are the elements of the $f$-th feature map. Different feature maps are created starting from different random initialisation of the weights and the biases. Non-linearity is introduced by further processing the $s_{m,n}^f$ elements by a proper activation function. A common and effective choice is the ReLU activation function:
\begin{equation}
    t_{m,n}^f = {\rm max}(0, s_{m,n}^f).
\end{equation}
At this point, each feature map is downscaled through a max pooling function, selecting the maximum $t_{m,n}^f$ every 2$\times$2 pool of feature map elements. The resulting maps are 1/4 the original size. The same convolution plus pooling procedure can be repeated several times starting from a given set of feature maps, until the fully connected (or dense) layer is reached. This layer combines all its elements ({\it neurons}) with all the elements of all the feature maps at the lowest resolution. The convolutional and pooling layers have the purpose of extracting the main features of the images, exploiting typical image properties like shift invariance and localised information (neighbouring pixels have usually related properties), and reduce dramatically the computational cost of the network, scaling down the number of pixels and limiting the number of parameters to be trained to the weights and the biases (in our case, 10 parameters per feature map). The dense layer correlates all identified features providing a compressed version of the input image. Each connection between the latent space neurons and each of the elements of the lowest resolution convolutional layer accounts for a trainable parameter (an additional weight) and, in general, sets the computational cost of the network. The dense layer is followed by an upsampling network, which reconstructs the full image through convolutional and un-pooling layers specular to the downsampling ones, starting from the feature map. The result is returned by the output layer. 



Besides trainable parameters (weights and biases), additional network parameters, whose values are used to control the learning process and cannot be estimated via training, are referred as {\it hyperparameters}. For our DDA, hyperparameters are:
\begin{itemize}
    \item the learning rate $\mu$, that regulates the step size of the iterative error minimisation procedure;
    \item the batch size $B$, that defines the number of images that are propagated through the network in one iteration step;
    \item the number of epochs $\epsilon$, i.e. the number of times a full dataset is processed;
    \item the tile size $T$, that defines the size of the tiles (see Section \ref{sec:data}).
\end{itemize}

The DDA has been implemented using the Keras software package (\citealt{chollet2015keras}), distributed as part of the Tensorflow framework (\citealt{tensorflow2015-whitepaper}), version 2.3.0. The GPU implementation of Tensorflow has been exploited to speed up the computation. We used distributed computing, based on the adoption of the MPI4Py library\footnote{https://mpi4py.readthedocs.io/en/stable/index.html}, in order to run in parallel various combinations of the hyperparameters of the model and identify the most accurate setup.

Training and tests have run on the Marconi100 (M100) High Performance Computing (HPC) system \footnote{https://www.hpc.cineca.it/hardware/marconi100} available at the CINECA Italian Supercomputing centre. M100 consists of 980 computing dual socket nodes with 2x16 cores IBM POWER9 AC922 at $2.6$ GHz and 256 GB memory per node. Each node is equipped also with four NVIDIA Volta V100 GPUs interconnected through Nvlink 2.0. The network is a Mellanox IB EDR DragonFly++ and 8 PB of GPFS storage are available in the scratch filesystem.  

\section{The Synthetic Observations}
\label{sec:data}

Synthetic images have been generated in order to be the closest possible to actual observations, leading to reliable training sets for the DDA. 

\subsection{Sky models}

We created maps of diffuse radio emission from recent cosmological MHD simulations, produced with the grid code {\it Enzo} \citep[][]{enzo14}, in which we evolved a uniform primordial magnetic seed field of $B_0=0.1 ~\rm nG$ (comoving) from $z_{\rm in} \geq 45$ to $z=0$, within a  $100^3 \rm ~Mpc^3$ volume simulated with  $2400^3$ dark matter particles and cells.
The uniform and constant spatial cell resolution is  $41.65 \rm ~kpc$ (comoving), while the constant mass resolution for dark matter particles is $m_{dm}=8.62 \cdot 10^{6} M_{\odot}$. 
This simulation is one of the largest MHD cosmological simulation ever run, and it has already applied by our group also to the study of X-ray emission from the cosmic web \citet[][]{va19}, as well as to detailed comparison with real low-frequency radio observations of possible diffuse synchrotron emission from the cosmic web, both using LOFAR \citep[][]{lo21L} and MWA \citep[][]{vern20}, and in combination with advanced models for generation of catalogs of star forming and radio galaxies \citep[][]{2021arXiv210607901H}. 

The radio emission from cosmic shocks at various redshifts has been calculated assuming that shocks can accelerate relativistic particles producing continuum and polarised radio emission \citep[e.g.][]{2011JApA...32..577B}.  The synchrotron emission model by \citet{hb07} has been assumed, which requires the jump condition of each cell undergoing shocks (computed from the simulation), the local value of the magnetic field and the electron acceleration efficiency as a function of Mach number (which is calibrated on shocks internal to galaxy clusters, as in \citealt{va15radio}). 

The cosmological model adopted in our simulations has the following parameters: $\Omega_\Lambda=0.692$, $\Omega_M=0.308$, $\Omega_b=0.0478$, $H=67.8 ~\rm km/s$ and $\sigma_8=0.815$. 

Images have been generated by stacking together 4 different snapshots of the simulation at increasing redshift volumes along the line of sight, with the same procedure described in \citet{2021MNRAS.500.5350V}.

Based on previous works we do not expect to detect a significant amount of radio emission from the cosmic web at very high redshift  \citep[e.g.][]{va15radio}.  Moreover, the recent analysis of the stacked signal in between hundreds of thousands of pairs of halos observed with MWA by \citet{vern20} has presented the possible detection of the radio signature from the cosmic web in such crowded environment, whose median redshift is $\langle z \rangle \approx 0.14$. For this reason, and in order to spare the computational time and memory needed to produce very long simulated lightcones \citep[e.g.][]{2021arXiv210607901H} we limited our analysis of simulated lightcones to  $z \approx 0.15$. A detailed description of the procedure adopted to generate mock radio lightcones can be found in \citet{2021MNRAS.500.5350V} and \citet[][]{2018MNRAS.480.3749G}.

A set of 1000 independent sky model images was generated by applying random rotations to each of the different redshift slices used to produce the lightcones. The nominal angular resolution of the images (pixel size) is 2 arcsec.



In summary, the described procedure was used to generate $2000 \times 2000$ pixel images, sampling a field of view of $1.1^\circ\times1.1^\circ$. These images are indicated as {\it Sky images} and represent the dataset against which the autoencoder is trained.

\subsection{Training datasets}
\label{sec:random}

Sky images are used to generate the synthetic observations by introducing noise and instrumental artefacts expected in a realistic scenario. For this, we have selected as reference instrument the LOFAR radio telescope (https://www.astron.nl/telescopes/lofar/) in its High Band Antenna (HBA) configuration. We have adopted the sensitivity of 0.1 mJy/beam taken as a conservative value characterising  the LOFAR {\it Two-Metre Sky Survey} (LoTSS, \citealt[][]{2019A&A...622A...1S}) at an angular resolution, defined as the full width half maximum (FWHM) of the synthesised beam, of 6 arcsec (corresponding to 3 pixels). We have investigated images with three different kinds of noise and artefacts, with increasing complexity. 

Pure random noise has been added to the Sky images as random (gaussian distributed) fluctuations with $0$ mean and variance equal to the LoTSS sensitivity. The dataset resulting from this process is referred as {\it Noise images}. Example of Noise images are shown in Figure \ref{fig:randomnoise7} and \ref{fig:randomnoise1}.



We have then applied LOFAR's instrumental response to our images, which includes the effects of the instrumental point spread function (PSF) and of the primary beam. Random noise is also added to visibilities at this stage.
The beams introduce different kind of artefacts compared to random noise and their treatment is specifically addressed by using the clean method, producing the deconvolved images that in the rest of the paper we will be referred as {\it Clean images}. The cleaning procedure, however, cannot fully remove the signatures of the beams. Residual artefacts influence the quality of the final image and potentially affect any attempt of automating the identification/classification of sources, which may be confused with the leftovers of the cleaning procedure. 

The Clean images have been generated as single frequency LOFAR HBA observation at $150$MHz of the matching Sky models. Visibilities corresponding to a 8 hours observation have been computed using the ``predict'' mode of the WSClean software (\citealt{offringa-wsclean-2014}, \citealt{offringa-wsclean-2017}). The effect of the HBA primary beam is introduced at this stage. 
Finally imaging is performed using once more the WSClean software, adopting Briggs' weighting, with robustness parameter equal to 0, and correcting for the primary beam. Deconvolution is carried out using the clean method. It must be stressed that the cleaning process has not been optimised for each single image, as it was applied to hundreds of samples by an automated procedure. Figure \ref{fig:cleanimages} shows an example of the restored deconvolved images.

The WSClean software produces also the {\it Dirty images}. These are the raw result of the imaging procedure, affected by all the possible noise and artefacts that then are mitigated through the cleaning algorithm. We have experimented the autoencoder also on this dataset, in order to verify its behaviour in these extreme conditions. Example of Dirty images are presented in Figure \ref{fig:dirtyimages}.

The $2000 \times 2000$ pixels images of the Sky, Noise, Clean and Dirty datasets are further divided into square tiles that becomes the actual training set of the network. Tile size is chosen in order to be the smallest possible still representative of the features to be identified (both sources and artefacts, in particular those due to the dirty beam, which can span large areas of the image). In this way we maximise the size of the training sets without losing the significance of each single input image, reducing, at the same time, the memory footprint of the DDA (see Section \ref{sec:performance} for details). An effective tile size for our images results to be 128$\times$128 pixels. The only relevant drawback of the tiling procedure is represented by possible small mismatches at the tile boundaries, which can impact the processed sources. We further discuss this issue in Section \ref{sec:discussion}.




\section{Data Denoising}
\label{sec:results}

We have studied the performance of the DDA for different kind of input datasets, starting from images with no noise, that allows us to investigate the impact of the numerical method on the results, progressively including random noise and artefacts produced by the aperture synthesis methodology adopted for radio interferometric data. The tests have the objective of $i$) verifying the effectiveness of the autoencoder in removing errors from the images, preserving the properties of the signal; $ii$) probe the capacity of the autoencoder to effectively perform the segmentation of the images, disentangling the sources, even in their faintest components, from the noisy background. 

The training and the testing of the autoencoder are implemented as follows:
\begin{enumerate}
\item the training program reads the input parameters, the hyperparameters and sets up the network;
\item training images are read from FITS files stored on disk;
\item the minimum of emissivity is set to a floor value of $10^{-8}$ Jy/arcsec$^2$ in order to reduce the dynamical range of the data. This threshold is less than 1/100 of the noise, so it does not lead to any loss of information. The logarithm is calculated; 
\item the results are normalised so that each image has values between 0 and 1;
\item images are divided into tiles;
\item tiles are serialised to feed the network;
\item mini-batches of tiles are offloaded to the GPU and there processed for the training; 
\item once reached convergence, the trained network is saved in a file. 
\item the trained network is loaded from a file by the evaluation program which performs the denoising of test images, selected from a dataset never ``seen'' before by the DDA. The processed image is finally saved in a FITS file.
\end{enumerate}

The adopted architecture of the autoencoder consists in the input and output layers, 2 convolutional plus 2 pooling layers for encoding, 2 convolutional with 2 unpooling layers for decoding and 1 dense layer made of 400 neurons for the Identity and 200 neurons for noisy data. The details of how such architecture has been selected are discussed in Appendix \ref{sec:appA}. Hyperparameters tuning is instead discussed in Appendix B.

\subsection{Identity}
\label{sec:identity}

We firstly focus on the ability of the DDA to exactly reproduce the input image. This allows us to verify the effectiveness of the method, with no influence of additional noise or artefacts. Starting from the discussion in Appendix \ref{sec:appB}, the adopted hyperparameters setup has been set further adopting the mean relative error ($\sigma_{MRE}$) measure:
\begin{equation}
    \sigma_{MRE} = {1\over N} \left [ \sum_{i=1}^N {\vert X_i - Y_i \vert\over X_i}\right ]_{{\rm flux}>10^{-7}},
\label{eq:mre}
\end{equation}
where $X_i$ and $Y_i$ are the flux densities in the Sky and encoded images respectively. The MRE metric has been calculated considering only the $N$ pixels with flux density bigger than $10^{-7}$Jy/arcsec$^2$. This in order to select only those pixels that are expected to contribute to detectable signals (such threshold corresponds to a conservative 1/30 fraction of the noise), excluding, at the same time, all the $\sim$ 0 flux pixels which form most of the Sky images and would dominate the MRE average, hiding the error related to the sources. The optimal set-up has proved to be: $\mu = 0.0001$, $B = 50$, $\epsilon = 100$, which gives an MRE = $0.159 \pm 0.039$.

\begin{figure*}
\begin{center}
\includegraphics[width=0.95\textwidth]{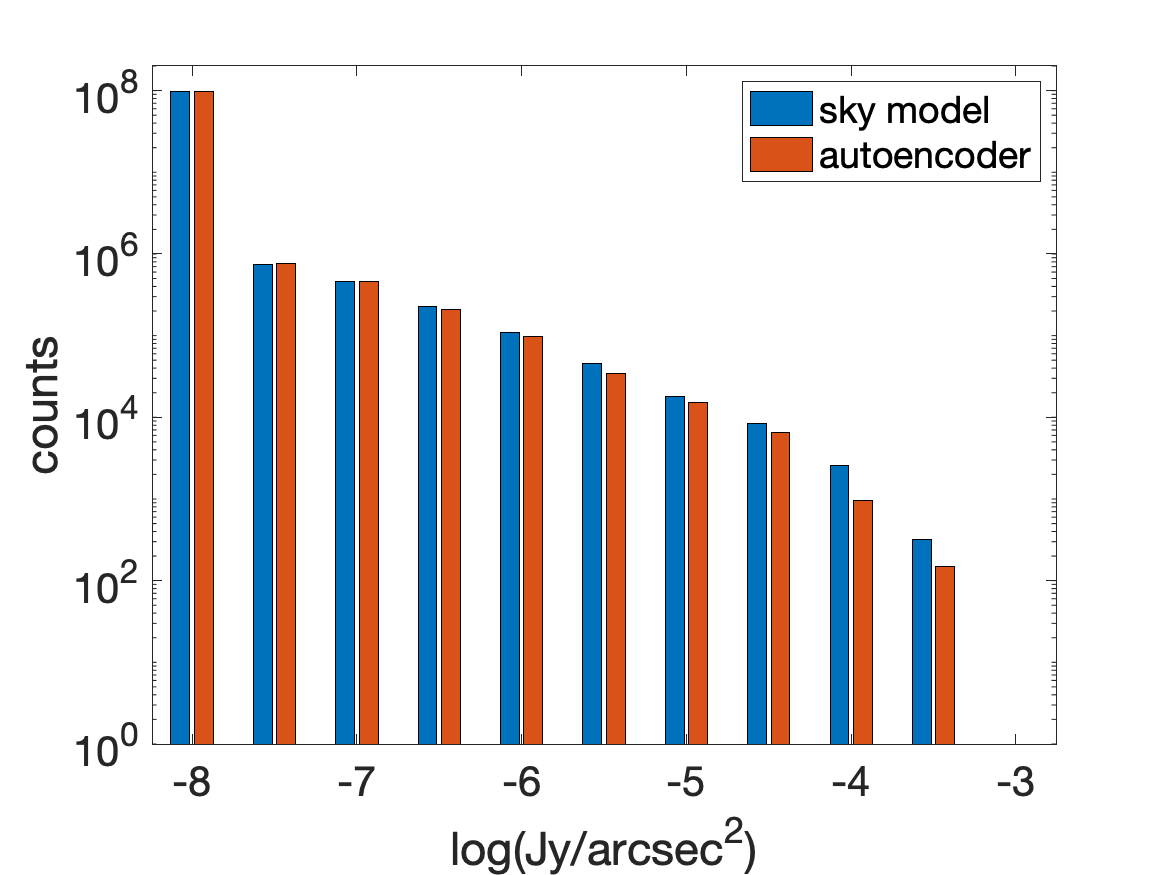}
\caption{Flux density distribution from Sky (blue bars) and DDA processed images (red bars).}
\label{fig:histo1}
\end{center}
\end{figure*}

The MRE shows that the encoded images differ on average by 15\% from the Sky ones. To further investigate such difference, we have compared in Figure \ref{fig:histo1} the flux density distributions of the input Sky images and of the encoded images. Although the flux density distribution of the encoded data closely follows that of the Sky data, an offset in the flux density scale of encoded images, towards lower value, is present. 

\begin{figure*}
\begin{center}
\includegraphics[width=0.48\textwidth]{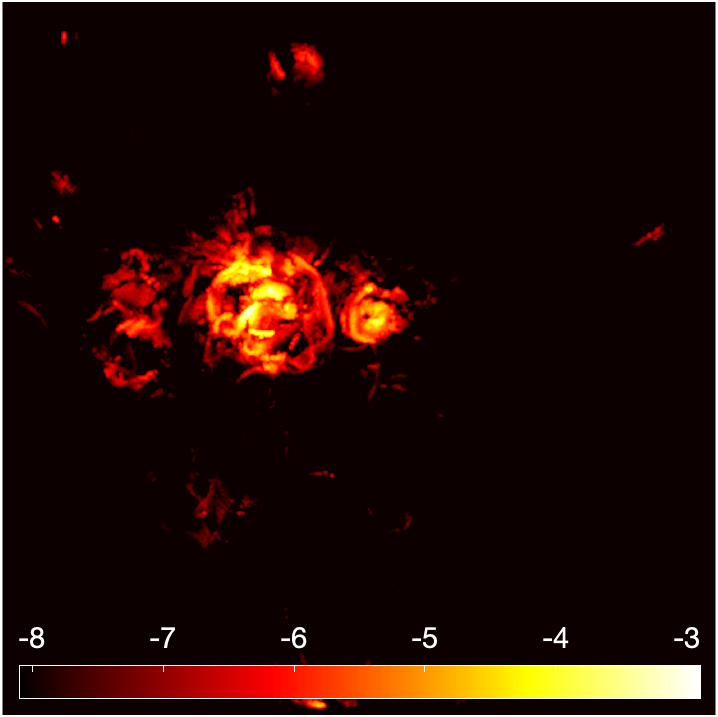}
\includegraphics[width=0.48\textwidth]{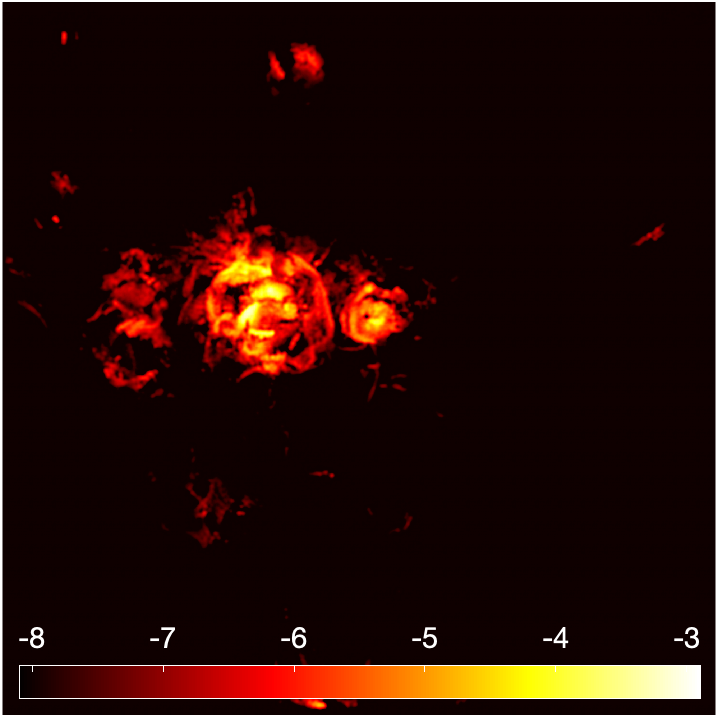}
\includegraphics[width=0.48\textwidth]{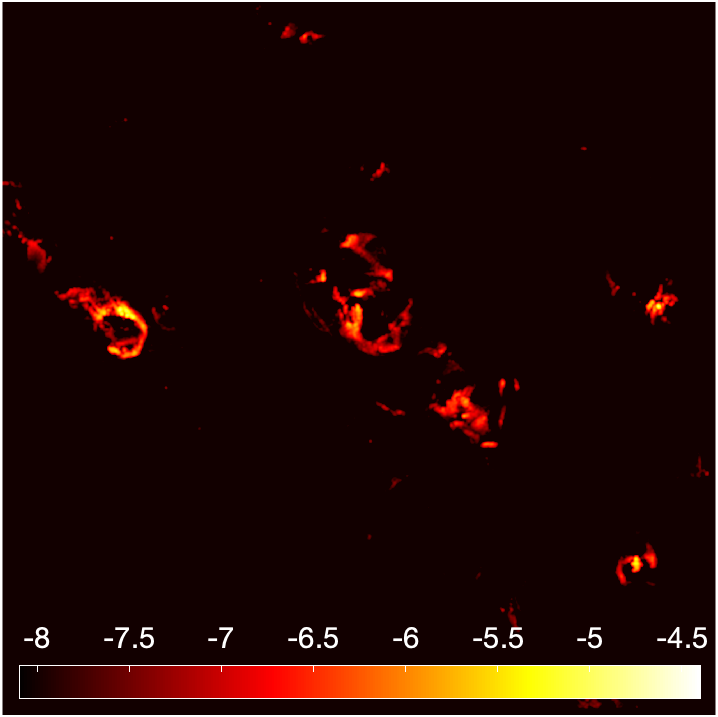}
\includegraphics[width=0.48\textwidth]{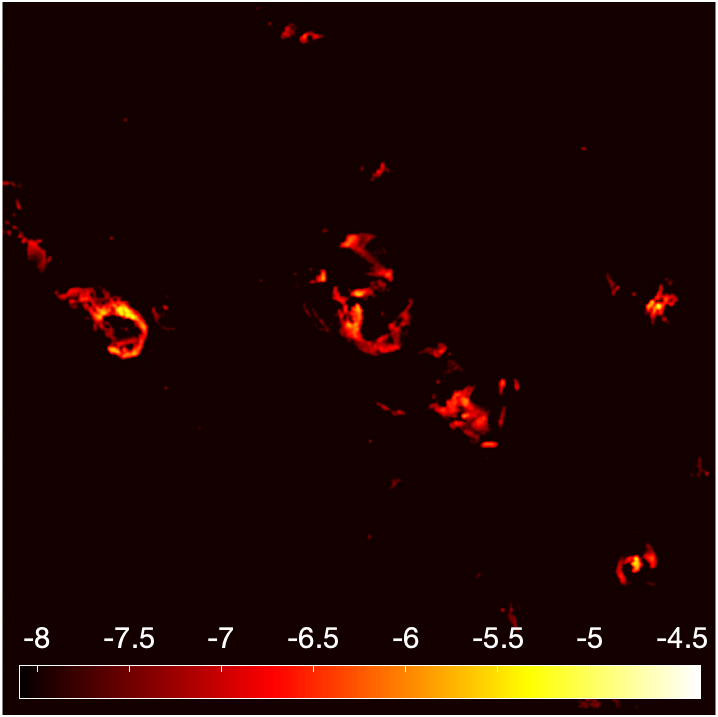}
\caption{Flux density maps of sky1 (top row) and sky2 (bottom row) cases, representing a 1.1 square degrees area of the sky. Noise-less Sky images are on the left while DDA encoded images are on the right. Flux densities are in log(Jy/arcsec$^2$) scale.}
\label{fig:compare1}
\end{center}
\end{figure*}

In order to explain such offset, we compare in Figure \ref{fig:compare1} the flux densities from two different cases. The first field (top row) is dominated by a large, $\sim  10^{15} ~M_\odot$ cluster of galaxies ({\it sky1} case). The flux densities reach values up to $10^{-3}$Jy/arcsec$^2$, i.e. well above the sensitivity of an instrument like LOFAR, and therefore most of the structures in the field is detectable. The second field (bottom row) features instead several smaller groups of galaxies (several $\sim 10^{13} ~M_\odot$), interconnected by filaments, and  is instead characterised by fainter diffuse sources ({\it sky2} case) with only a  few pixels above $10^{-5}$Jy/arcsec$^2$. 

\begin{figure*}
\begin{center}
\includegraphics[width=0.48\textwidth]{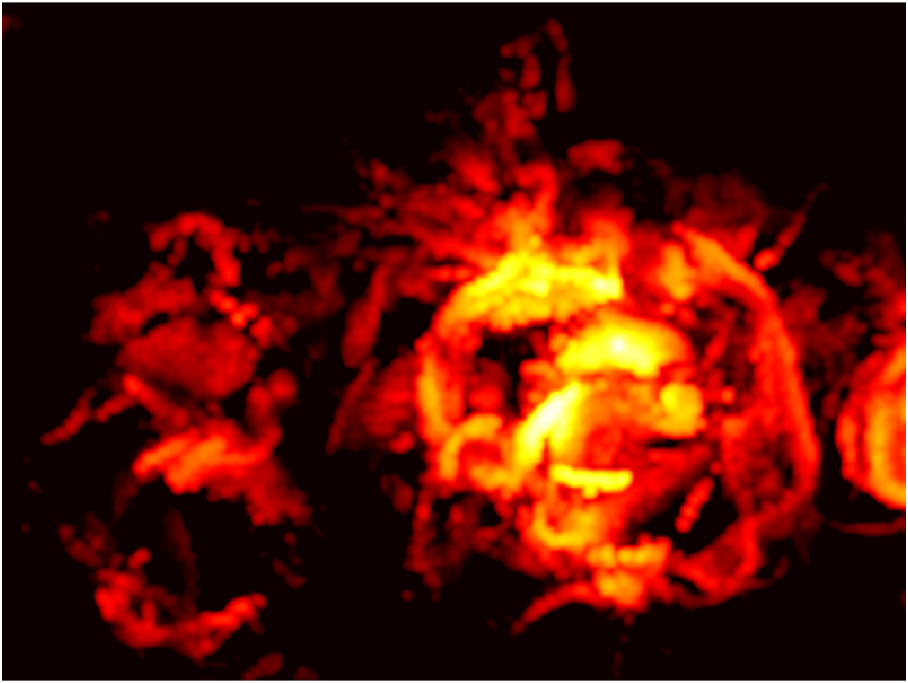}
\includegraphics[width=0.48\textwidth]{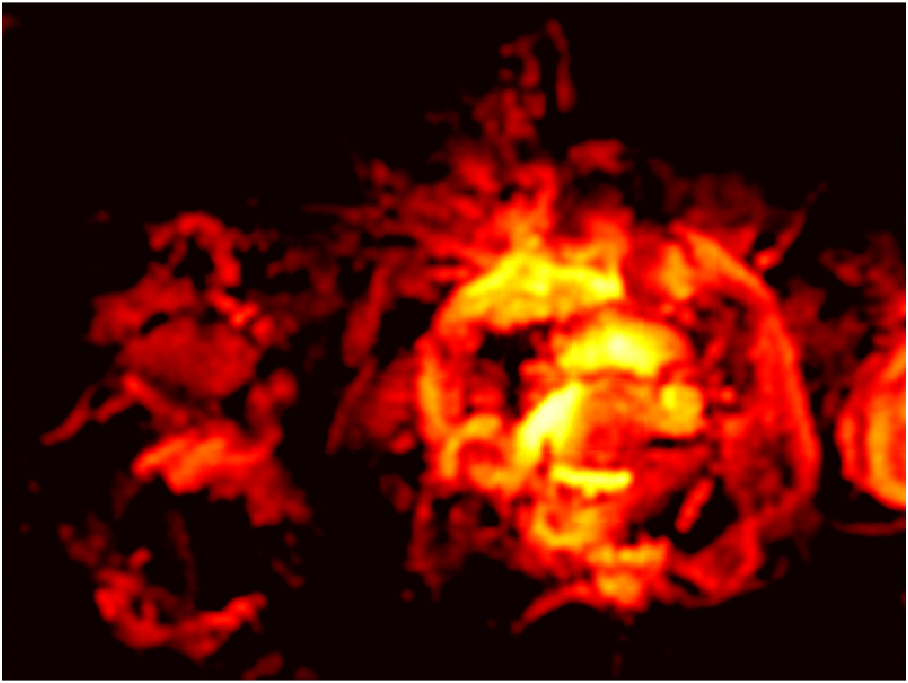}
\caption{Zoom on the main source in the sky1 model highlighting the slight blur of the encoded image (right panel) compared to the original Sky image (left panel).}
\label{fig:zoom1}
\end{center}
\end{figure*}

The two fields are extremes in the continuous spectrum of cosmic structures that radio surveys target, and of the current challenges in detecting both internal radio shock waves powered by merger events, as well as dimmer and more peripheral accretion patterns associated with the acceleration of radio emitting electrons in the rarefied conditions of strong accretion shocks. Both cases  will be used throughout the paper as reference cases. Sky images are on the left, DDA encoded ones on the right. From a visual inspection, it is clear how the encoded images are very similar to the Sky ones. 
However, with a more detailed analysis, by zooming on one of the sources of sky1 and presented in Figure \ref{fig:zoom1}, we can notice that encoded objects are slightly more diffused (or blurred) than the Sky ones, leading to the generalised decrease of their surface brightness, as observed in the histograms. This leads to the estimated absolute error we noticed above, and it can be interpreted as a consequence of using the convolutional kernel in the encoding layers, together with the decrease of resolution in the deep layers. Both steps result into a blurring numerical effect on the steepest flux density gradients in the input data. This is discussed in more details in Appendix\ref{sec:appA}.

\subsection{Random Noise}
\label{sec:noisetests}

We have added random noise to the Sky images as described in Section \ref{sec:random}, investigating how this is treated by the DDA. The hyperparameters setup is $\mu = 0.0001$, $B = 50$, $\epsilon = 100$ (see Appendix B for details). For comparison, we have also performed the same analysis using a Gaussian based and a FFT based denoising algorithms. The former performs a Gaussian convolution to the data, mimicking the application of a synthesised beam shaped kernel to enhance source detection (\citealt{1933RSPTA.231..289N}). The value of the standard deviation $\sigma$ of the Gaussian adopted for the filtering is equal to three pixels (6 arcsec), i.e. approximately twice the width of the synthesised beam. The latter is a standard filtering in the Fourier space, which works by setting the high frequency modes of the Fourier Transform of the image to zero. The two are implemented using the SciPy gaussian\_filter and fftpack modules (https://docs.scipy.org/doc/).

\begin{figure*}
\begin{center}
\includegraphics[width=0.95\textwidth]{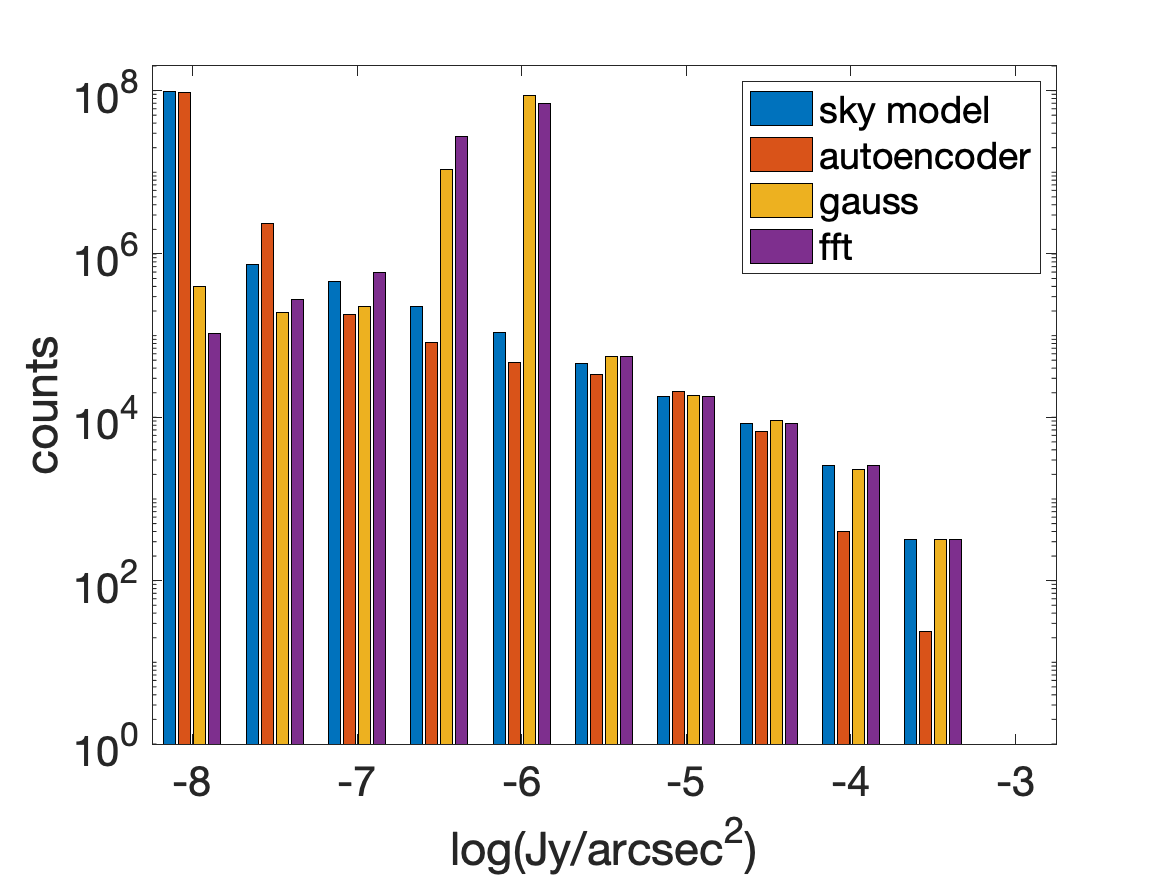}
\caption{Flux density distributions from the Sky, DDA denoised, gaussian and FFT filtered images in the case of pure random noise.}
\label{fig:histo-randomnoise}
\end{center}
\end{figure*}

In Figure \ref{fig:histo-randomnoise} we show the flux density distribution comparing the Sky model, the images denoised by the autoencoder and those denoised by the Gaussian and FFT filters. The most relevant features in the histograms are that 1) as already highlighted in the previous section, the autoencoder tends to blur the highest fluxes, by redistributing their flux density to a slightly wider area, 2) at signal to noise ratio $S/N \leq 1$, the autoencoder closely follows the sky distribution, while the other two methods are strongly affected by the noise. Throughout the paper, S/N is defined as the ratio between the pixel flux density in $\rm Jy/arcsec^2$, at the scale of the imaging resolution and the instrumental sensitivity of $=0.01 ~\rm mJy/beam$. The noise at the resolution scale is the most significant for the present study, targeting the detection of barely detectable signals.

\begin{figure*}
\begin{center}
\includegraphics[width=0.48\textwidth]{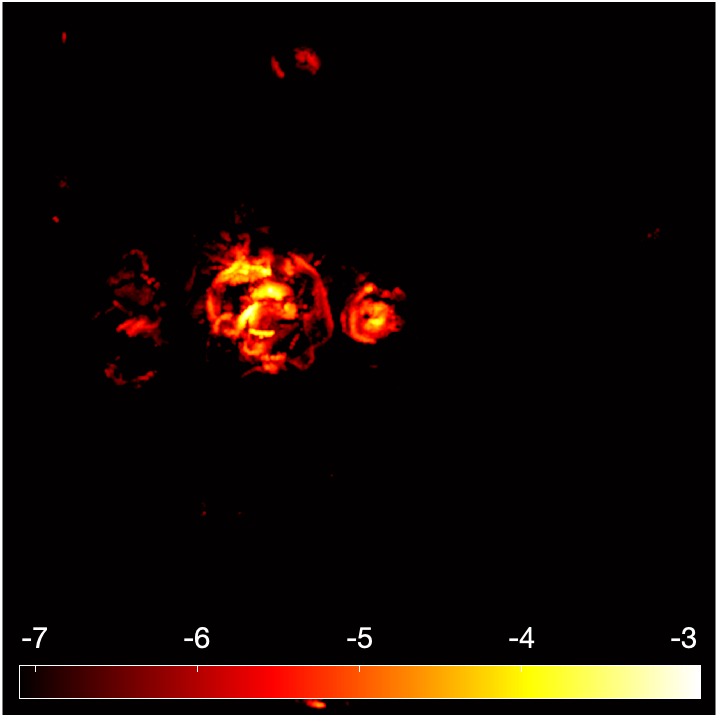}
\includegraphics[width=0.48\textwidth]{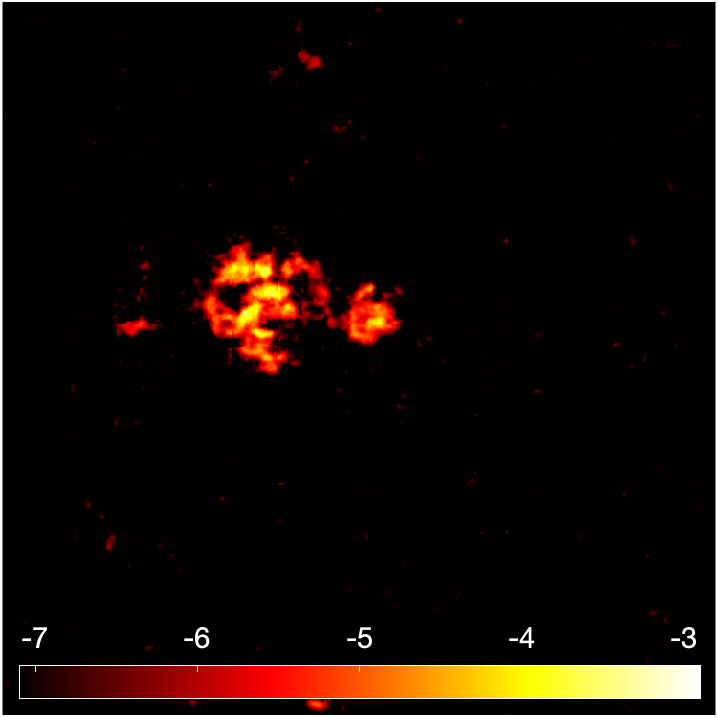}
\includegraphics[width=0.48\textwidth]{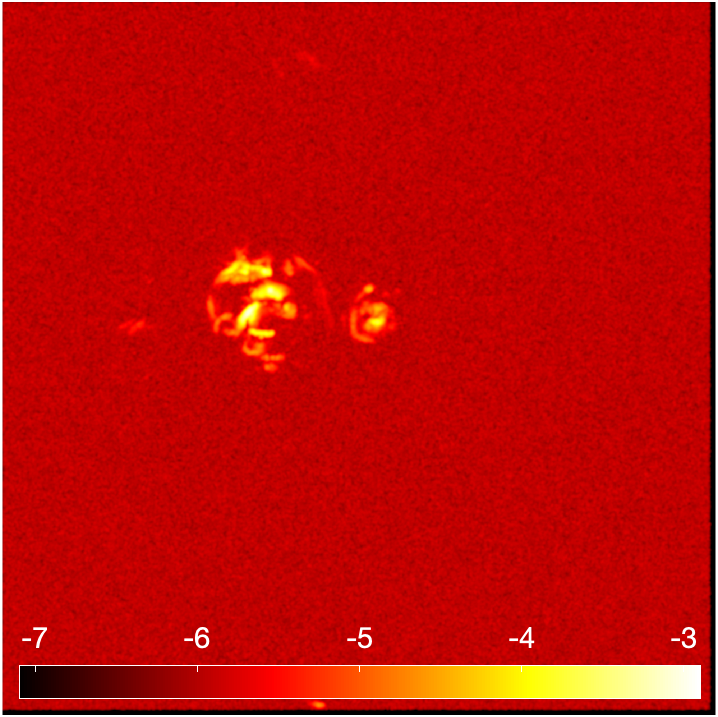}
\includegraphics[width=0.48\textwidth]{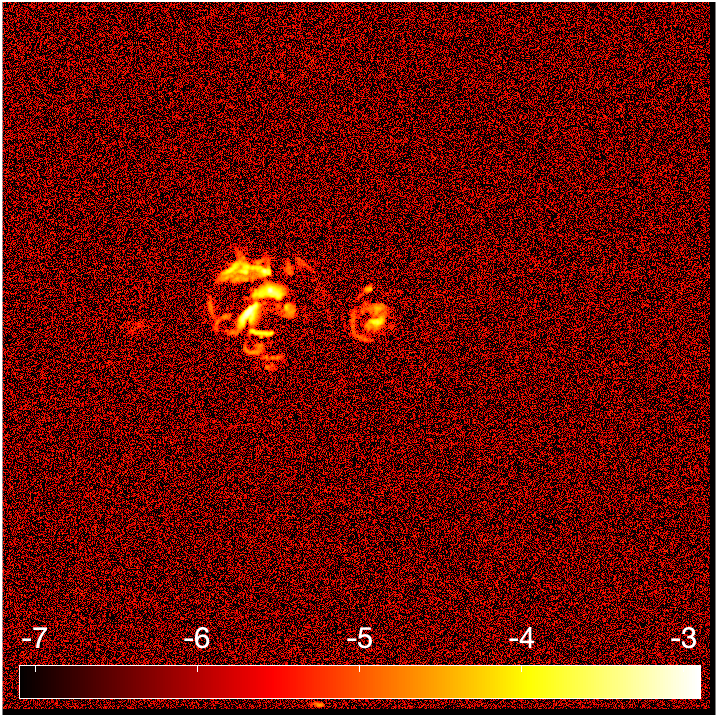}
\caption{Flux density maps of the sky1 model comparing the Sky reference image (top left), the autoencoder denoised image (top right), the gaussian filtered image (bottom left) and the Noise input image (bottom right). Flux densities are in log(Jy/arcsec$^2$) scale.}
\label{fig:randomnoise7}
\end{center}
\end{figure*}

\begin{figure*}
\begin{center}
\includegraphics[width=0.48\textwidth]{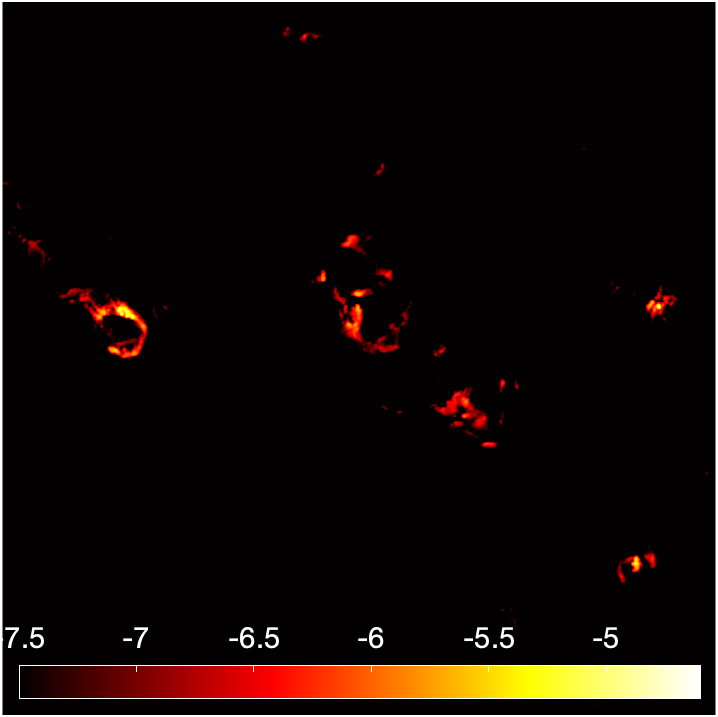}
\includegraphics[width=0.48\textwidth]{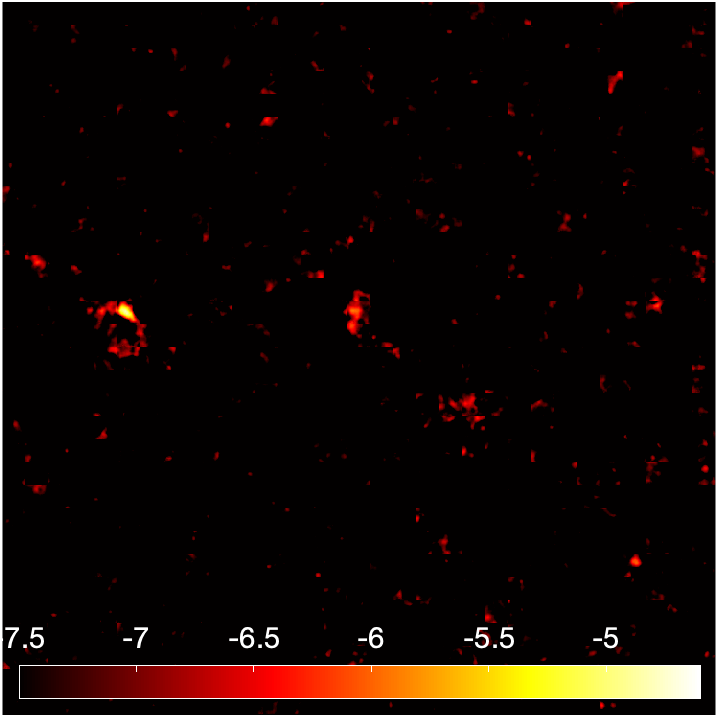}
\includegraphics[width=0.48\textwidth]{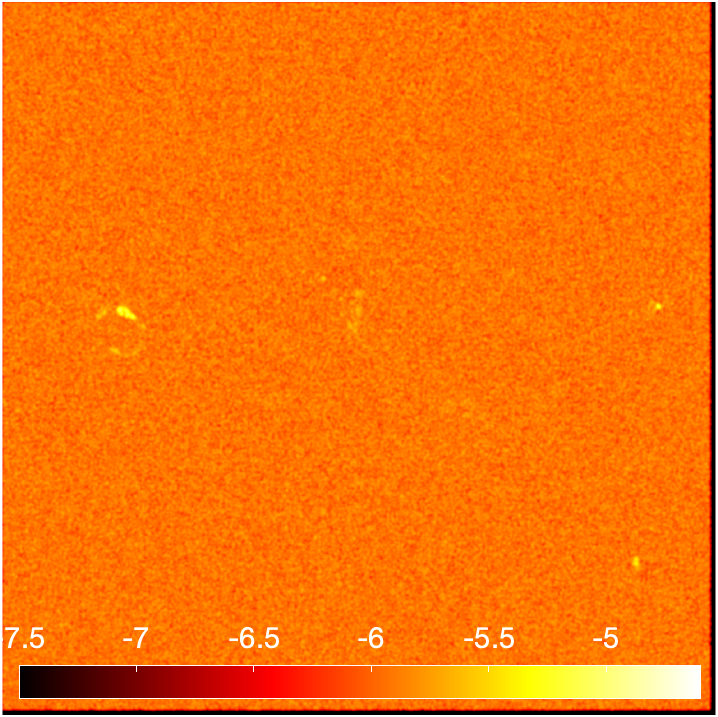}
\includegraphics[width=0.48\textwidth]{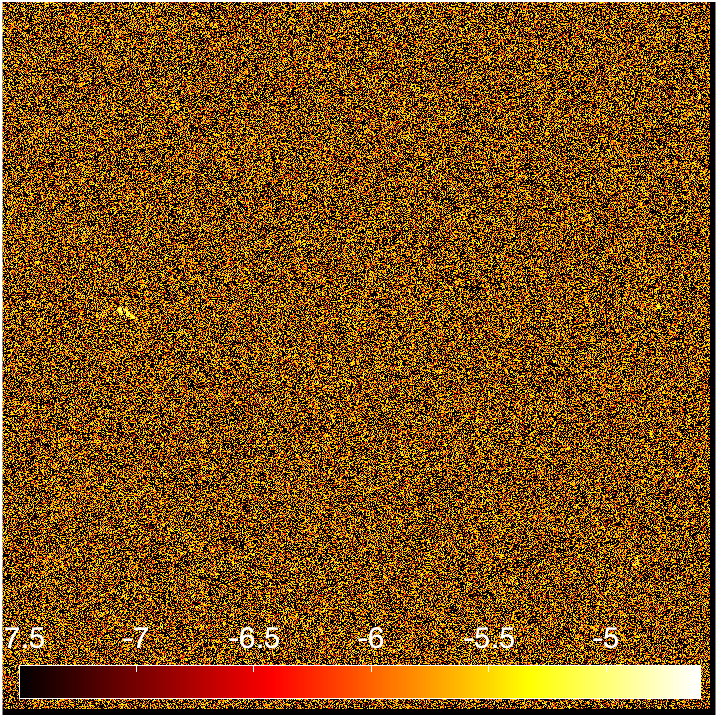}
\caption{Flux density maps of the sky2 model comparing the Sky reference image (top left), the autoencoder denoised image (top right), the gaussian filtered image (bottom left) and the Noise input image (bottom right). Flux densities are in log(Jy/arcsec$^2$) scale.}
\label{fig:randomnoise1}
\end{center}
\end{figure*}

In Figure \ref{fig:randomnoise7} and \ref{fig:randomnoise1} the flux density distributions of models sky1 and sky2, respectively, are shown. The Sky image is presented on the top-left panel and can be compared to the results of the autoencoder (top-right) and of the Gaussian filter (bottom-left). The bottom-right Figure show the input noisy image. The results of FFT based denoising are similar to those of the Gaussian filter (not shown). In all images, we used a threshold of $10^{-7}$Jy/arcsec$^2$, corresponding to a signal to noise ratio of 1/30.

\begin{figure*}
\begin{center}
\includegraphics[width=0.48\textwidth]{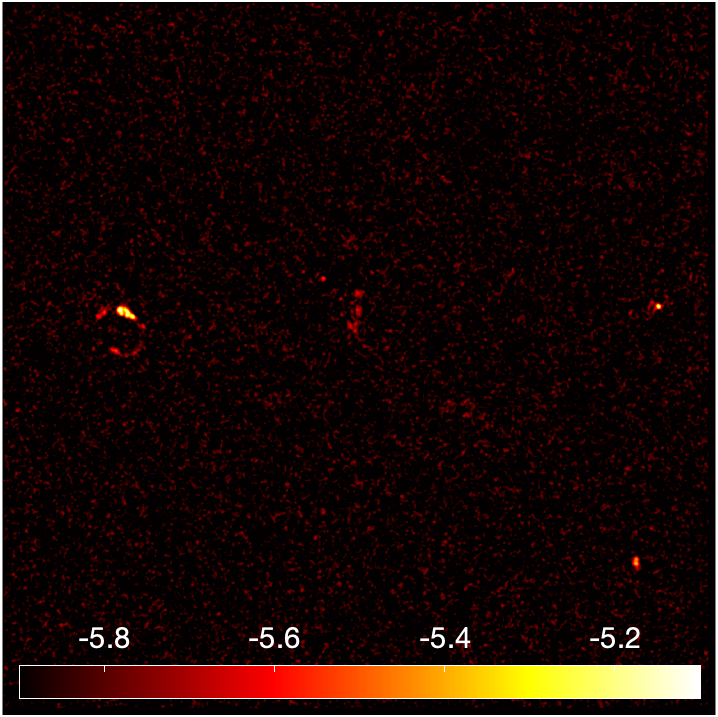}
\includegraphics[width=0.48\textwidth]{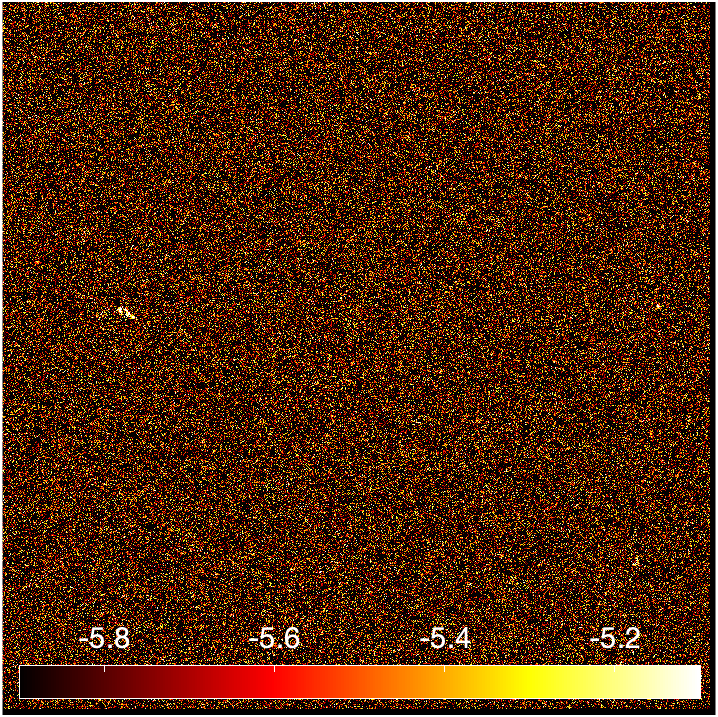}
\caption{Flux density maps of the sky2 model comparing the gaussian filtered image (left panel) and the Noise input image (right panel) with an applied threshold at S/N = 1/3 (=$10^{-6}$ Jy/arcsec$^2$). Flux densities are in log(Jy/arcsec$^2$) scale.}
\label{fig:randomnoise1-2}
\end{center}
\end{figure*}

In the images processed by the autoencoder, the random noise has been largely suppressed, while diffused sources are well traced and  preserve their original morphology. As pointed out above, the brightest peaks tend to be fainter and more diffused than in the Sky model. This is a result of the blurring and diffusive effect of the multi-resolution convolutional algorithm and is particularly evident for the sky1 model, whose emission is well above the noise, so that the source can be clearly identified even in the noisy image. The Gaussian cleaning smooths out the noise, making its removal easier, with a thresholding filter. Things becomes more challenging when fainter sources are considered, as for model sky2. Here, sources are barely visible in the noisy image as they have a signal to noise ratio $\leq 1$. The encoded image can identify most of the sources and their main features are reproduced, although faintest spots are wiped out. Some spurious artefacts also appear, which can be confused with weak sources. However, such artefacts are random and can be easily removed by stacking several DDA encoding of the same image. In the Gaussian filtered image, signals are still confused in the smoothed noise. However, setting a flux density threshold at S/N = 1/3, we get the result shown in Figure \ref{fig:randomnoise1-2}, where high frequency noise is fully removed but only the brightest peaks of the flux density distribution are present. 


We have adopted the cross-correlation analysis to perform a quantitative estimate of the correspondence between images, hence of the ability of the DDA to perform an accurate segmentation of the sources confused in the noisy image, disentangling signals even at S/N $<$ 1. Cross-correlation is commonly used in signal processing to measure the similarity of two signals as a function of the displacement of one relative to the other. For 2D $N\times M$ pixels images $A$ and $B$, the normalised correlation matrix $C$ is defined as:
\begin{equation}
    C(k,l) = {1\over NM}\sum_{j=0}^{N-1} \sum_{i=0}^{M-1} {(A(i,j)-\bar A)(B(i+k,j+l)-\bar B) \over 
               \sigma_A \sigma_B},
\end{equation}
where $\bar A$ and $\bar B$ are the mean values of the two images and $\sigma_A$ and $\sigma_B$ are their standard deviation. The indices of the correlation matrix represents the shift (displacement) of the two images. The correlation is normalised by the standard deviation of each quantity in order to allow for a direct comparison of different quantities. The normalised cross-correlation function $C_r$ is calculated as the average of $C(k,l)$ over elements having the same radial separation $r = (k^2 + l^2)^{1/2}$.  The $C_r$ function takes values between -1 and 1, the latter representing perfect linear correlation between quantities ($A\propto B$). The value -1, represents perfect linear anti-correlation. 
We recently widely explored the application of this technique as a tool to detect the gas component of the cosmic web in existing and future multi-wavelength surveys \citep[][]{gv20}, inspired by the latest applications to low-frequency radio observations \citep[][]{vern17}.

\begin{figure*}
\begin{center}
\includegraphics[width=0.48\textwidth]{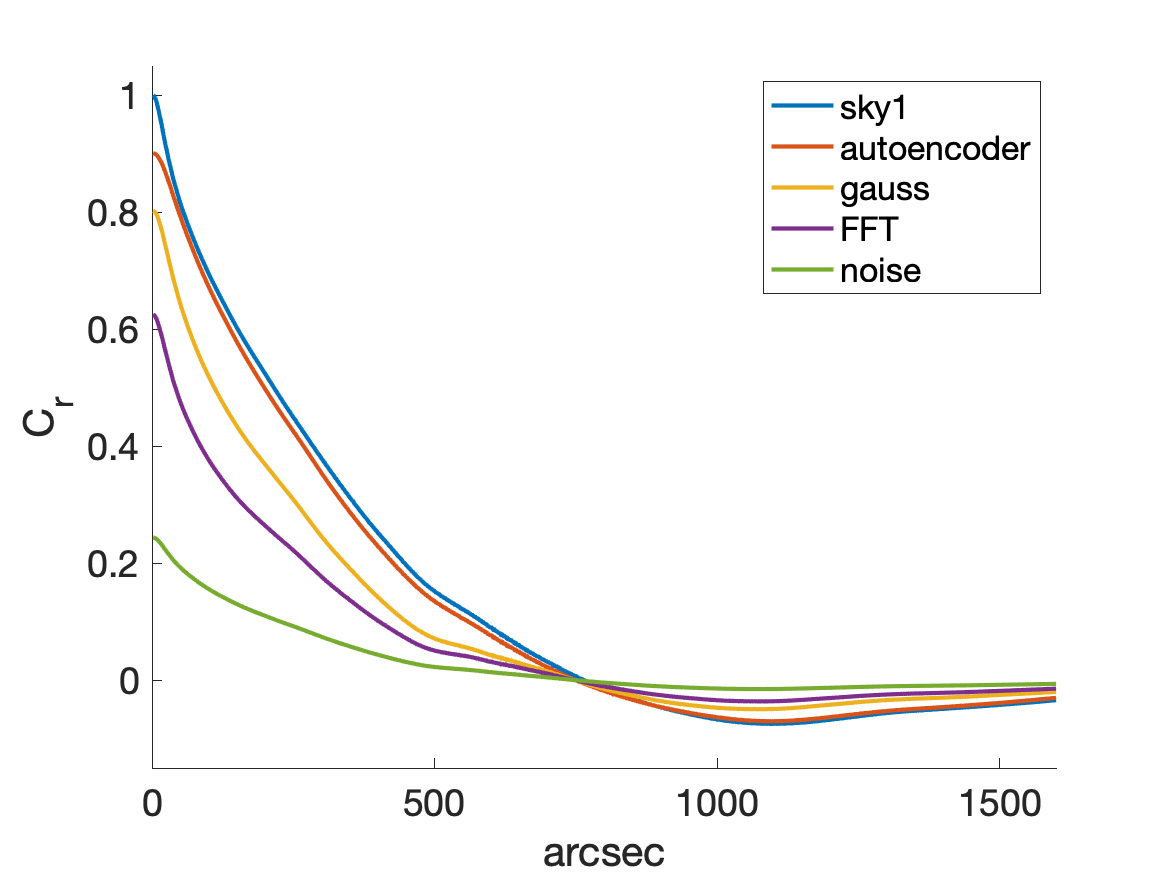}
\includegraphics[width=0.48\textwidth]{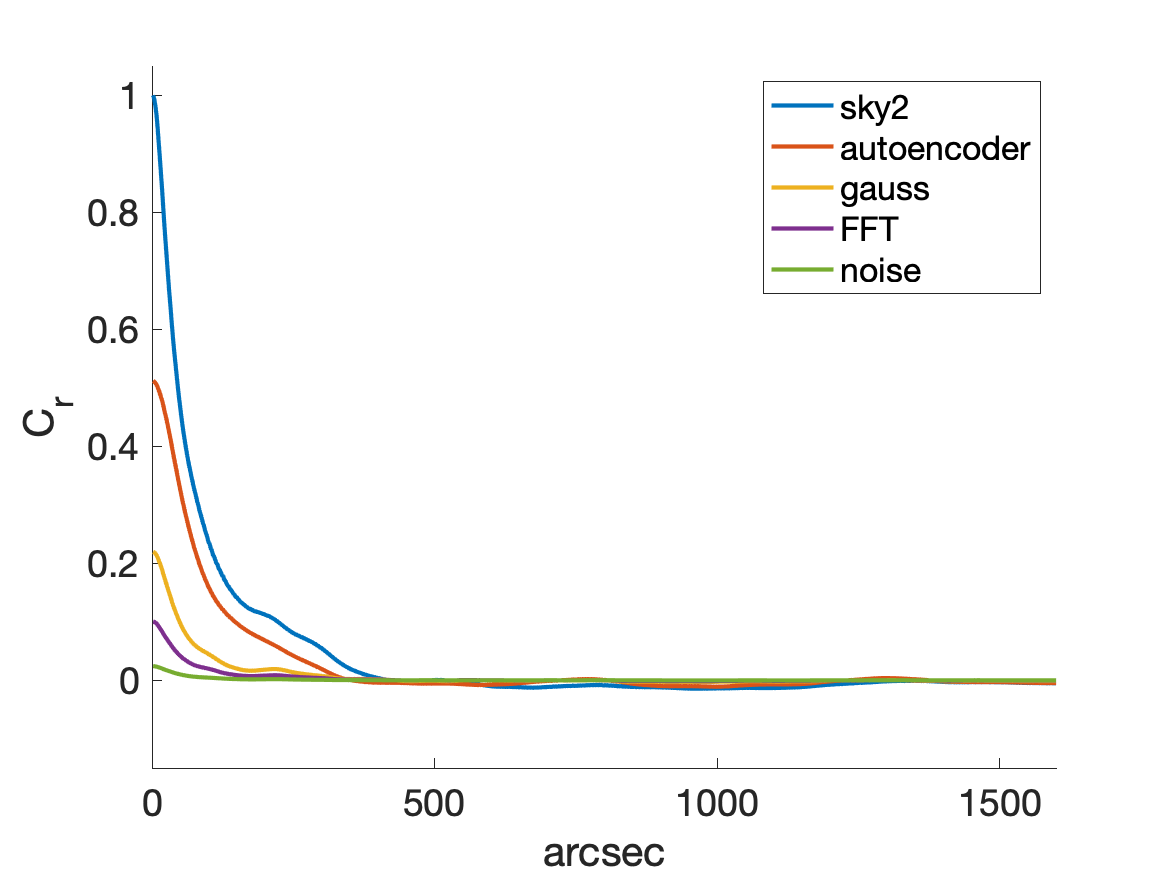}
\caption{Cross-correlation curves for the sky1 (left panel) and sky2 (right panels) Noise models. The blue curve is the autocorrelation of the Sky image, as a reference. The other curves show the cross-correlation of the DDA solution (red curve), of the gaussian and FFT filtered images (yellow and purple curves respectively) and of the noisy input image with the Sky image.}
\label{fig:crossnoise}
\end{center}
\end{figure*}

In Figure \ref{fig:crossnoise} we show the cross-correlation for the sky1 and sky2 models. As a reference, the auto-correlation of the Sky image is shown (blue line). Cross-correlation of the encoded, the Gaussian and the FFT filtered images with the Sky models have been calculated. Furthermore, the cross-correlation of the Noise images with the Sky images are shown. This last curve (green line), shows how strongly the noise affects the Sky images, especially in the sky2 case, where the correlation is almost completely lost. All applied filters are able to improve the correlation, with the autoencoder best resembling the Sky curve in both cases. For sky1, where S/N is higher and the sources are extended (auto-correlation is 0 at 800 arcsec), all three methods reproduce appropriately the original image. However, only the autoencoder is able to reproduce the Sky model until very large separations, i.e. at the lowest values of the signal. The match gets slightly worse for $\leq 20$ arcsec separations, due to the discussed impact of numerical blurring. The sky2 model is the most challenging, having a low S/N and smaller sources, as shown by the auto-correlation function, which drops to 0 already from separations of $\sim 400$ arcsec. This is clearly shown by all curves, with correlations lower than for sky1. The autoencoder follows the trend of the auto-correlation, indicating that the shape of the sources is closely reproduced although with smaller flux densities. Therefore, the autoencoder appears to be highly effective in identifying the shape of the sources, tracing also regions at very low S/N ratio, like outskirts of diffused objects, which are crucial for the study of cosmic magnetism in the outer regions of large scale structures.  

\subsection{Clean and Dirty Images}
\label{sec:cleantests}

Clean an Dirty images are generated as described in section \ref{sec:random}. 
For data processing, the DDA hyperparameters have been set to $\mu = 0.0001$, $B = 50$, $\epsilon = 500$ (see Appendix B for details). In Figure \ref{fig:cleanimages}, the Clean and the DDA flux density maps of models sky1 and sky2 are presented. Firstly, we notice that the sky1 and the sky2 Clean images differ substantially. The effect of the beam is dominant for bright sources, as for the sky1 model, while random noise prevails for the faint sky2 model. In the first model, the autoencoder is capable of removing almost all artefacts and sources are all properly identified. Some spurious noisy spots are instead present in the sky1 image. They can be removed with a thresholding filter set at $10^{-7}$Jy/arcsec$^2$, which however, cuts out at the same time also the  sources outskirts. Alternatively, a stacking based procedure of multiple encodings of the same image can erase random artefacts, without affecting the sources. In the sky2 image, the faintest parts of the sources cannot be recovered as they get entirely wiped away in the Clean image, by the combined effect of the noise and the sidelobes (the ring-like artefacts generated by the telescope beams).

\begin{figure*}
\begin{center}
\includegraphics[width=0.48\textwidth]{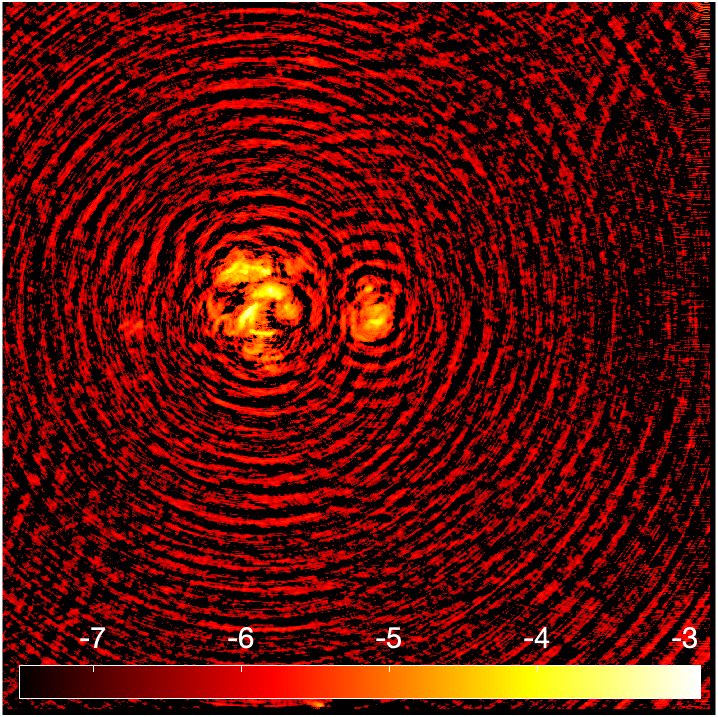}
\includegraphics[width=0.48\textwidth]{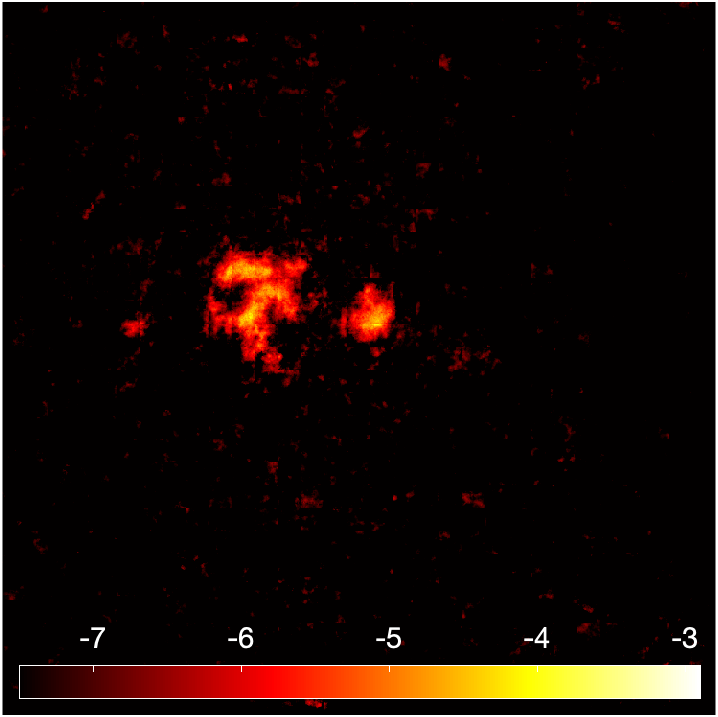}
\includegraphics[width=0.48\textwidth]{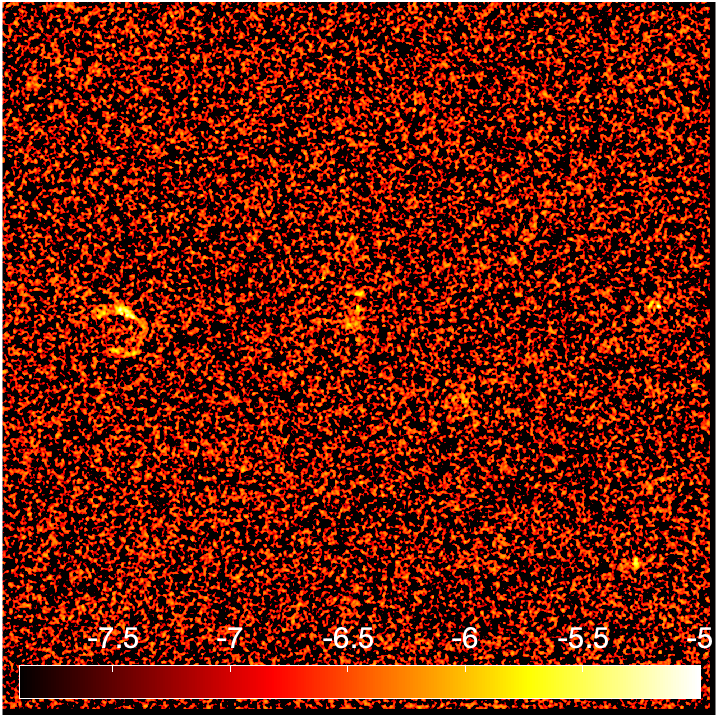}
\includegraphics[width=0.48\textwidth]{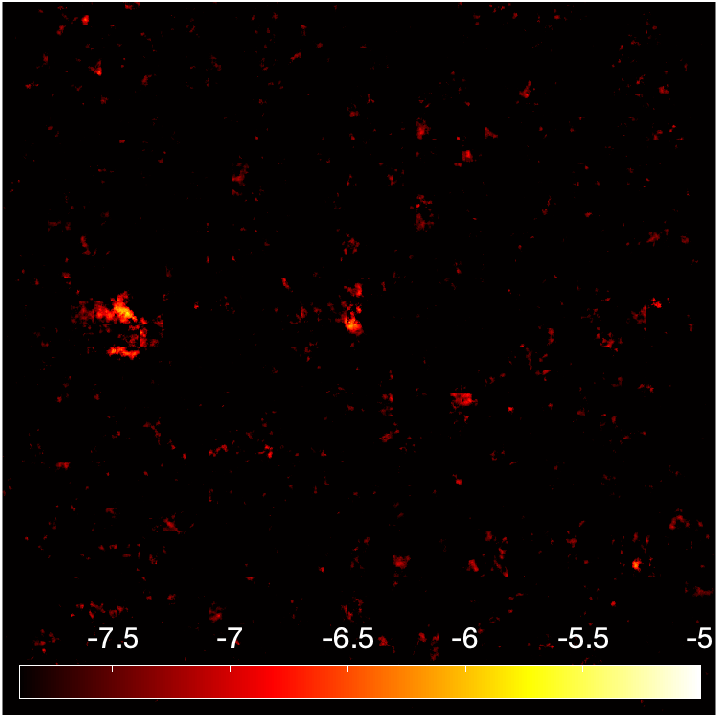}
\caption{Flux density maps of the sky1 (top row) and sky2 (bottom row) models comparing the Clean input image (left panels) to the corresponding DDA denoised image (right panels). Flux densities are in log(Jy/arcsec$^2$) scale.}
\label{fig:cleanimages}
\end{center}
\end{figure*}

\begin{figure*}
\begin{center}
\includegraphics[width=0.48\textwidth]{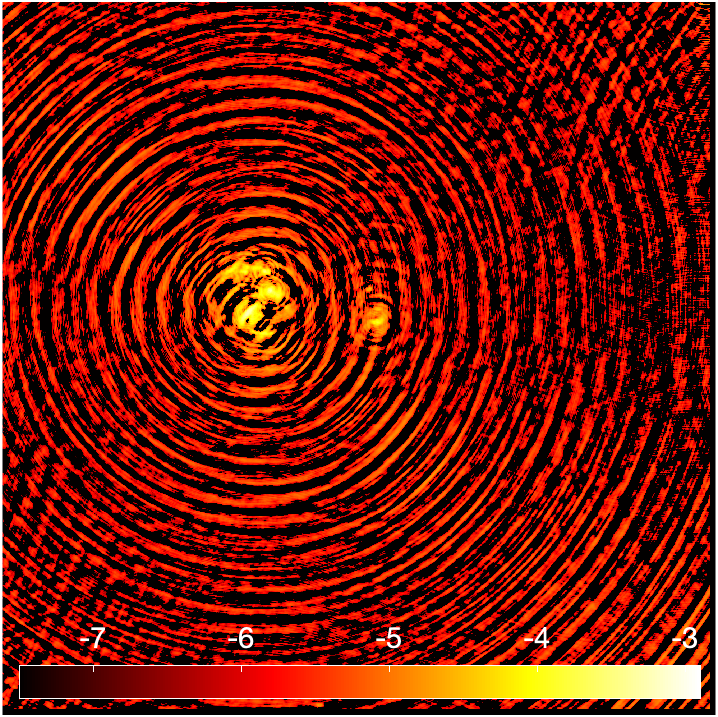}
\includegraphics[width=0.48\textwidth]{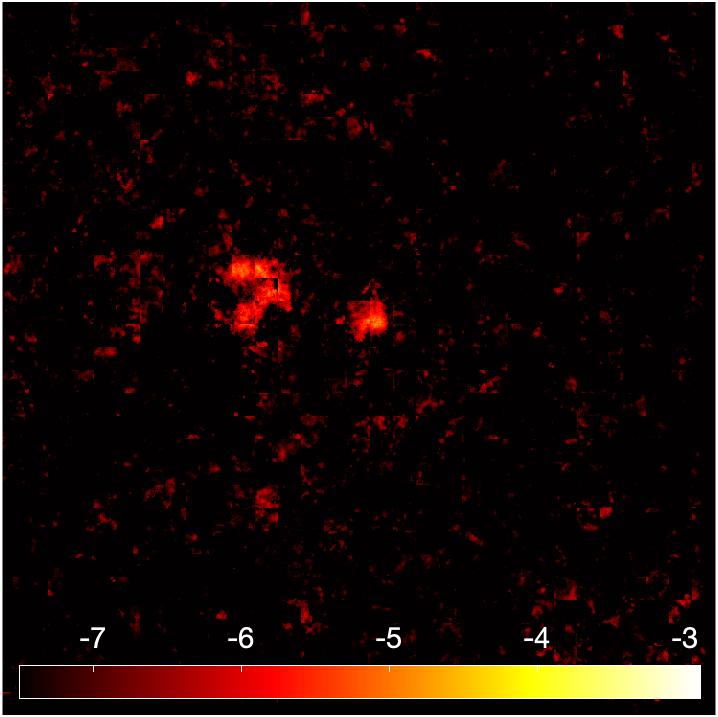}
\includegraphics[width=0.48\textwidth]{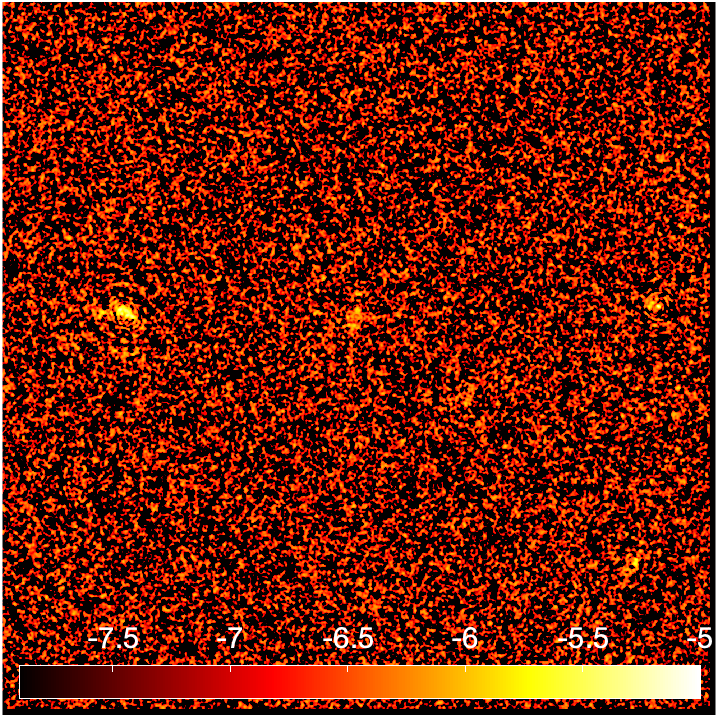}
\includegraphics[width=0.48\textwidth]{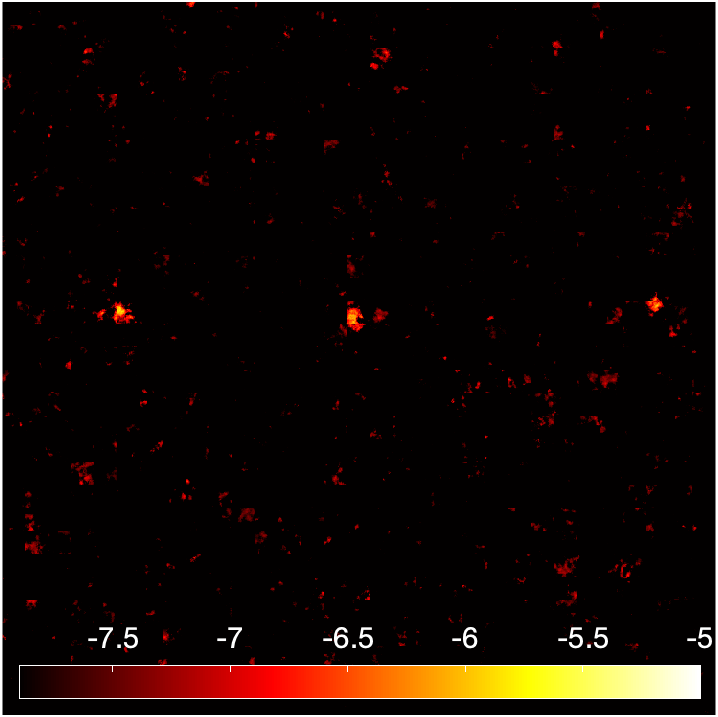}
\caption{Flux density maps of the sky1 (top row) and sky2 (bottom row) models comparing the Dirty input image (left panels) to the corresponding DDA denoised image (right panels). Flux densities are in log(Jy/arcsec$^2$) scale}
\label{fig:dirtyimages}
\end{center}
\end{figure*}

In the Dirty images, the effect of the beam is stronger, in particular for the sky1 model, as it is shown in Figure \ref{fig:dirtyimages}. Accordingly, the resulting denoised images show only a partial and dimmed reconstruction of the original sources, with spurious artefacts appearing. In the sky1 image an extended structure can still be detected, however the brightest peaks are all lost. Conversely, the sky2 image preserve only the highest peaks of the flux density distribution with a very limited amount of diffused matter traced. 

\begin{figure*}
\begin{center}
\includegraphics[width=0.48\textwidth]{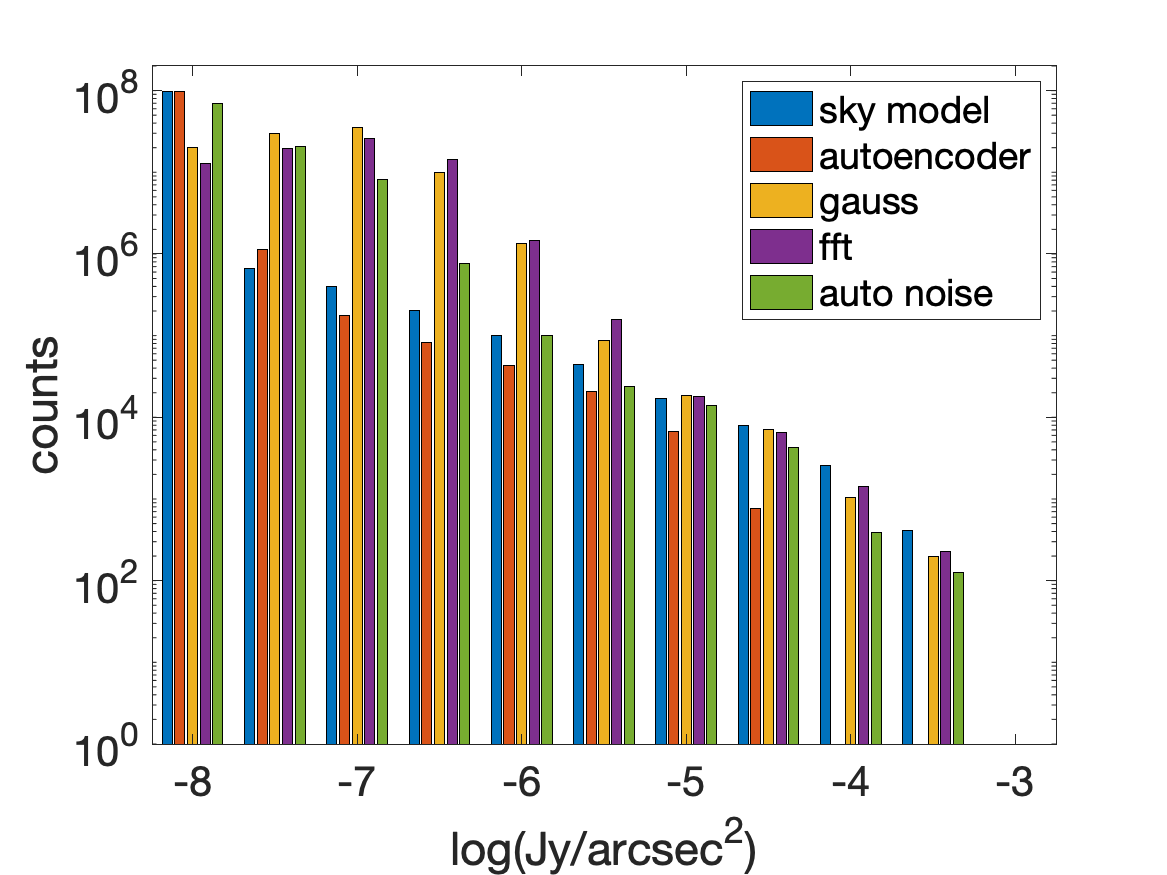}
\includegraphics[width=0.48\textwidth]{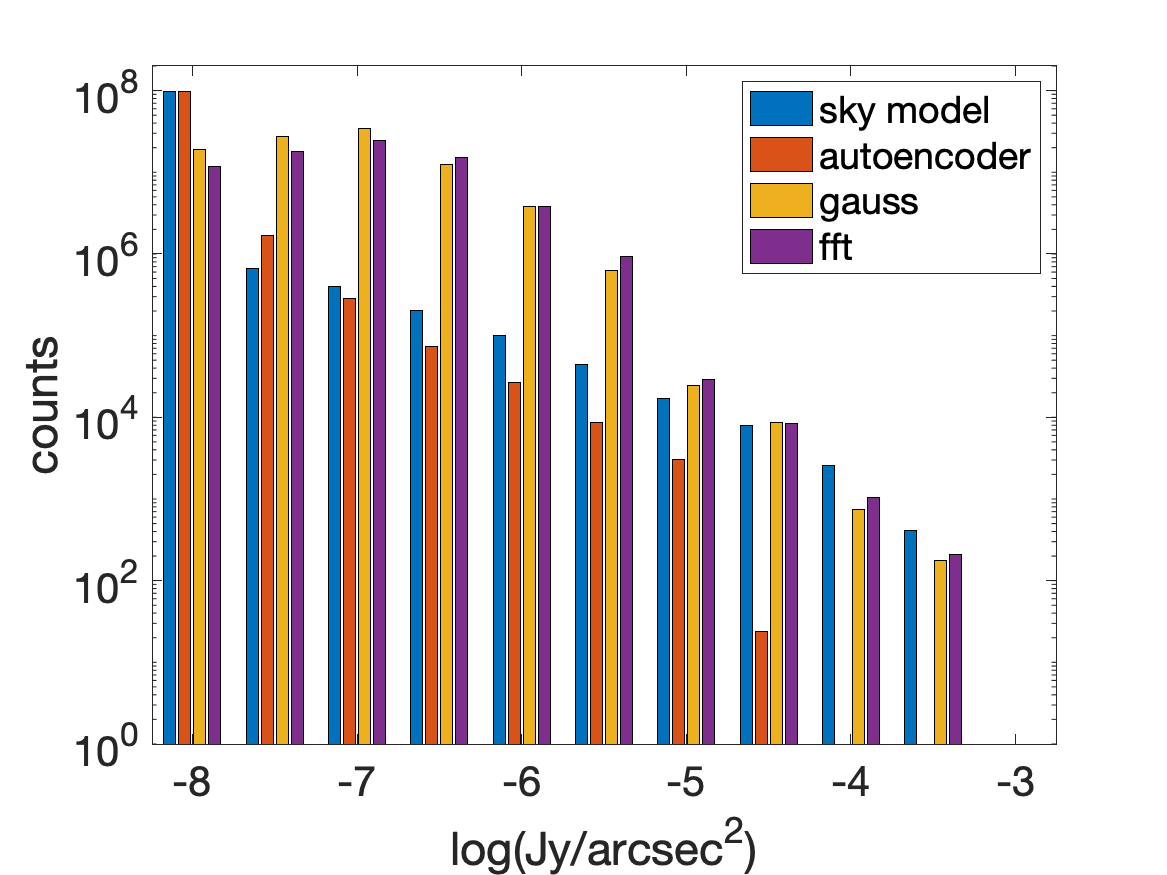}
\caption{Flux density distribution from Sky, autoencoder denoised, gaussian and FFT filtered images for the Clean dataset (left panel) and the Dirty dataset (right panel). For the Clean case also the Flux density distribution of Clean data denoised with the autoencoder trained with the Noise dataset is presented}
\label{fig:histoclean}
\end{center}
\end{figure*}

Although the autoencoder is capable of effectively removing noise and artefacts, a significant reduction of the flux density is found in all cases. This effect is confirmed by the flux density distribution presented in Figure \ref{fig:histoclean}, where a comparison also with the gaussian and FFT filters is presented. The Clean and the Dirty datasets are presented in the left and right panels respectively. It is clear how, for the autoencoder, high energy contributions are missing, their energy diffused at the lower energy bins. This effect is stronger, as expected, for the Dirty images. However, the autoencoder gets closer and closer to the correct energy distribution below $10^{-5}$Jy/arcsec$^2$, where other methods differ more from the Sky models. 

The dimming of the highest energy pixels was already found for the pure random noise data. However its impact on the results was smaller. We may then think to apply the network trained for random noise data to the images with fainter sources, that are dominated by the noise. It must be noticed that the generalisation of a model trained on a given dataset to data belonging to a different (although ``similar'') family, is known to be a challenge for any Machine Learning approach.

The ``auto noise'' bar of the histogram presented in Figure \ref{fig:histoclean}, shows the flux density distribution of such test. Higher energy  values are present with a flux density distribution closer to the Sky one for the highest bins. However, meaningful differences appears below $10^{-6}$Jy/arcsec$^2$, affecting the source components more challenging to detect, at the level or below the noise. 

\begin{figure*}
\begin{center}
\includegraphics[width=0.48\textwidth]{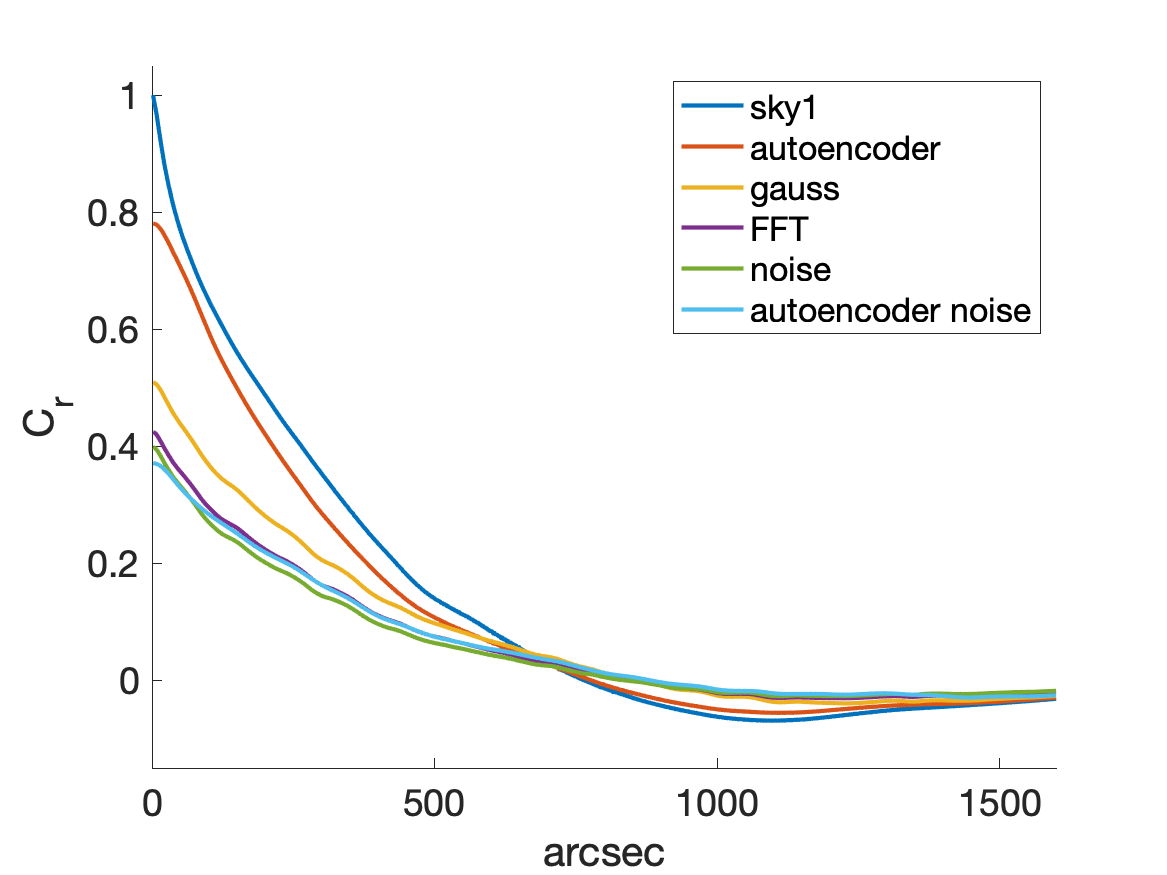}
\includegraphics[width=0.48\textwidth]{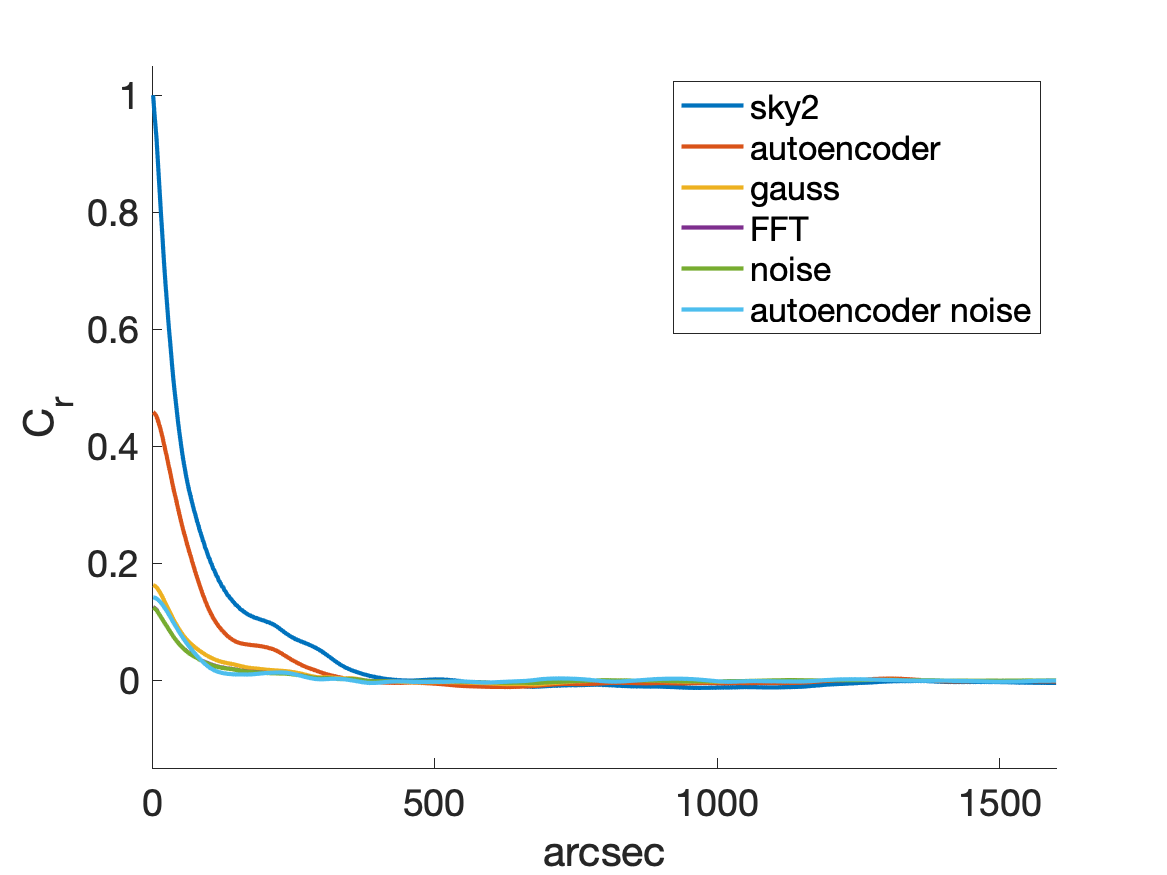}
\includegraphics[width=0.48\textwidth]{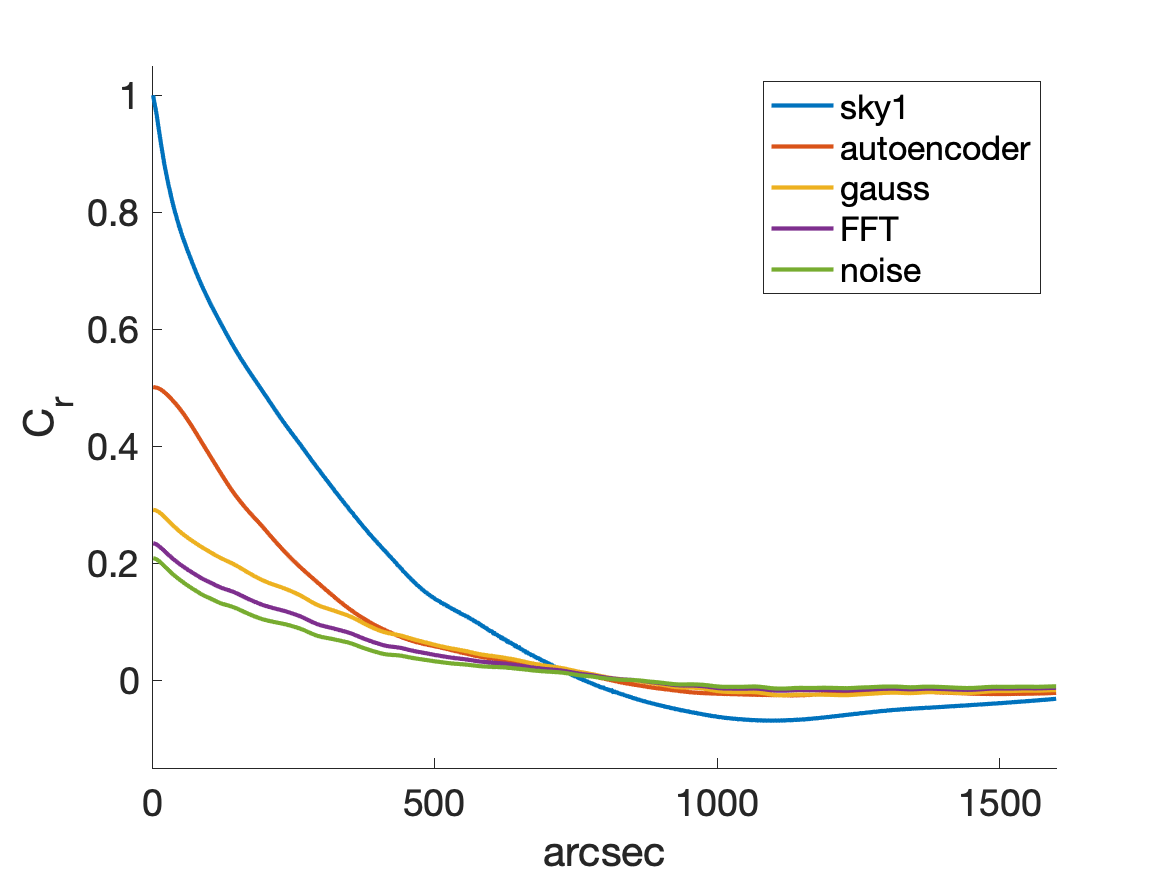}
\includegraphics[width=0.48\textwidth]{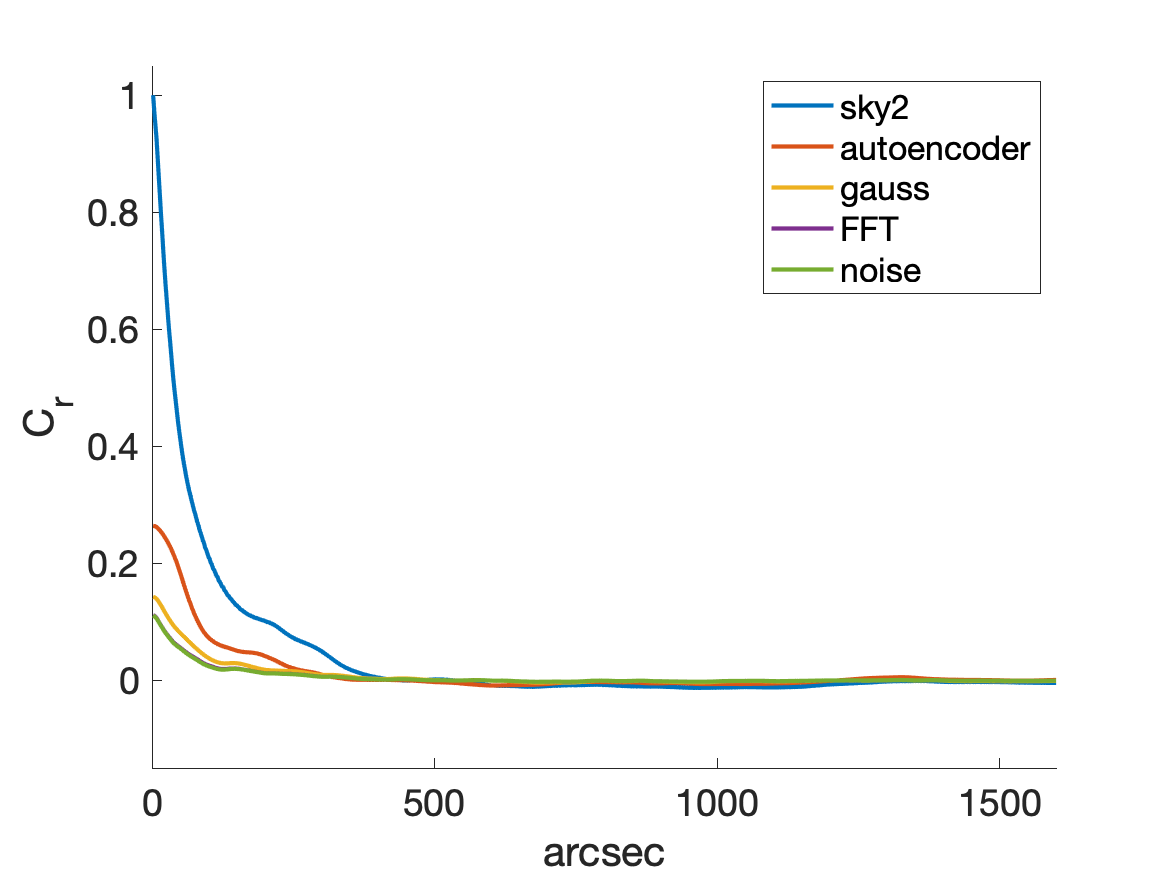}
\caption{Cross-correlation curves for the sky1 (left panel) and sky2 (right panels) Clean (top row) and Dirty (bottom row) models. The blue curve is the autocorrelation of the Sky image, as a reference. The other curves show the cross-correlation of the DDA solution (red curve), of the gaussian and FFT filtered images (yellow and purple curves respectively) and of the noisy input image with the Sky image. For the Clean data also the cross-correlation of the Clean images denoised with the autoencoder trained with the Noise dataset is presented (cyan curve).}
\label{fig:crossclean}
\end{center}
\end{figure*}

The cross-correlation analysis for the Clean dataset produces the results presented in the top row of Figure \ref{fig:crossclean}, where also the results obtained using the denoiser trained with the Noise images is calculated (``autoencoder noise'', cyan line). The autoencoder trained on the Clean data (red lines) follows the auto-correlation reference curve, for both the sky1 and the sky2 model. The sky2 cross-correlation is very similar to that obtained for pure noisy data. The result for sky1 is slightly worse than that obtained in the Noise case, but the degree of correlation is still remarkable. Hence, despite the dimming, the autoencoder is effective in tracing the morphology of even the faintest sources. The cross-correlation curves obtained adopting the denoiser trained with the Noise dataset to the Clean images has completely different behaviours in the sky1 and sky2 cases. The former is close to that produced by the Clean images trained denoiser, confirming the good correspondence between the Sky and the denoised image. For this kind of data, generalisation is successful. The second instead shows a worse performance, comparable to that obtained with the gaussian filter.

The bottom row of Figure \ref{fig:crossclean} shows the cross-correlation for the Dirty dataset. Correlation is lower than for Clean images, however a significant signal is still present even at large displacement ($\approx$200 arcsec) at least for the sky1 model. The sky2 model, as also highlighted by the maps, can only identify the highest flux densities of the distribution. 

\section{Computational Performance}
\label{sec:performance}

The computational requirements of the autoencoder, in terms of memory usage and computing time, are determined by its architecture, as well as by the image and the training dataset sizes. 

The memory usage of the DDA is controlled by the size of the fully connected layer combined with the size of the two deepest convolutional layers (one in the downscale and one in the upscale sweeps). Each element of the latter is connected to each element of the dense layer and the number of associated weights, as well as the required memory, scales proportionally to the product of the number of such elements in the three layers. The memory request also depends on the size of the input image, which sets the number of elements of each convolutional layer. Other parameters, like the size of the receptive window, the number of hidden layers or the number of feature maps have a minor memory footprint.

In the adopted configuration, the autoencoder network accounts for around 26 million trainable parameters (twice this value for the Identity case, given the larger size of the dense layer), corresponding to a memory requirement of 210 MB of data. The input image size (one tile) is 128$\times$128 pixels, each pixel being a 8 bytes quantity. This leads to additional 16 MB to store the elements of the various layers. 

In the training phase, the autoencoder loads all necessary data for the training into memory, resulting in a memory usage that varies from 6 to 12 GB, corresponding to around 25000 to 50000 tiles, depending on the dataset and the hyperparameter settings.

The computing time of the training phases (the most time consuming part) depends on the specific adopted set-up, in particular, by the number of epochs. The wall-clock time scales linearly with such parameter: the time to process a $\sim$ 25000 images training set is 1634 seconds for 100 epochs, 4824 seconds for 300 epochs and 8192 seconds for 500 epochs. The average time to process a single 128$\times$128 pixels tile is $\approx$ 512 microseconds, or 0.131 seconds per $N_{\rm image}$ =2000$\times$2000 pixels image. The remaining hyperparameters do not significantly affect the computing time.

The time required to train the network depends linearly also from the training set size: doubling the size of the input dataset and repeating the 100 epochs test, we get a computing time of 3205 seconds.

Finally the size of the single tiles impacts once more linearly the training time. For instance. 256$\times$256 pixels tiles require around 2 milliseconds to be processed, 4 times the time needed to process the 128$\times$128 pixels images. 

Additional time is necessary to load data in memory from FITS files and to perform the preprocessing of the images, as described in Section \ref{sec:results}. The preprocessing time is proportional to the training set size and corresponds to 440 seconds for 490 FITS files ($\approx$ 0.90 sec per file), which is the biggest training set that can be loaded due to memory constraints. Each FITS file is a 2000$\times$2000 pixels double precision image.

Once trained, the network is used to process the test images. 
The time to denoise our prototype 2000$\times$2000 pixels image is 0.31 seconds. This is approximately 3 times higher than the corresponding time obtained during the training. In fact, in order to increase the denoiser accuracy, the image is split in overlapping tiles whose number results a factor of 3 higher than in the training stage (where no overlap is necessary). 


\section{Discussion}
\label{sec:discussion}

The adoption of the convolutional deep denoising autoencoder Machine Learning approach to radio interferometric data has been investigated in details performing a number of tests to several different classes of input images, encompassing ideal Sky models, aiming at highlighting the impact of the numerical approach on the data, Noise data, addressing the performance of the denoiser for data affected by gaussian random noise, and Clean and Dirty data, in which data are additionally impacted by the aperture synthesis processing procedure.

We have optimised the network in terms of both its architecture and of the hyperparameters set-up. The former resulted in a network made of an input and an output layer, two convolutional plus two pooling layers for encoding, a specular configuration for decoding and one central dense, fully connected layer. Deeper architectures have proved to be less accurate, leading to an increasing (with depth) blur of the resulting images. Hyperparameters are challenging to tune, since they depend on the class of input images and they span large range of values that have to be combined in order to explore all the different possibilities. Furthermore, the stochastic nature of the algorithm introduces some additional variability to the accuracy and to the convergence of the process. The adoption of a parallel implementation of the denoiser, capable of running different hyperparameters setups and, to some extent, architectural choices (e.g. the number of neurons in the dense layer) at the same time on different computing nodes/GPUs, accelerated the exploration of this large parameters space in a dramatic way. The tuning remains, anyway, an outstanding task to optimally accomplish, requiring large computational resources and plenty of time and effort.

As metrics of the autoencoder performance, we have adopted the spatial distribution of the flux densities (the maps), the spectral distribution of the flux densities and the cross-correlation among the processed and the Sky maps. 

The maps show how the autoencoder is highly effective in removing the random noise, identifying the faint, extended outskirts of diffused structures. When the aperture synthesis process is included and the input image is affected by the response of an interferometer, the denoising becomes much more challenging. However, applied to the Clean data it is still capable of reconstructing images in which sources are still fully identifiable. With Dirty images, the method returns only a partial reconstruction of the original Sky models. 

We have verified that a simple thresholding filter is not capable of achieving any significant denoising on the data (apart from the extraction of the very highest peaks of the distribution). We have then compared the results of the autoencoder to those obtained with other denoising methods, based on gaussian smoothing or FFT filtering. These methods are effective at a level of a S/N ratio bigger than 1. The autoencoder is instead capable of effectively denoising sources with S/N at the scale of the imaging resolution of the order of 0.1 or even smaller. 

On the other hand, the maps show how the brightest peaks of the flux density distribution in the images denoised by the autoencoder are dimmed as a consequence of blurring, which smooth the energy of the brightest peaks, redistributing it on the neighbouring pixels. Blurring results to be an effect of the numerical diffusion due to the multi-layer convolution and pooling processes. It has been quantified by performing a flux density distribution (spectral) analysis, which gives an estimate of the reliability of the denoised flux densities. The spectral analysis shows how in fact highest energies are shifted to lower bins and how this effect increases with the complexity of the image to denoise. On the other hand, the flux density distributions are always consistent with those of the Sky data at S/N ratios $<$ 10, with a good correspondence at S/N $<$ 1, where the gaussian and FFT filters cannot reproduce the correct spectrum. 

As previously noticed, small artefacts can be introduced in the encoded images by the tiling procedure, due to mismatches at the boundaries of neighbouring tiles. This issue has been minimised (although not completely solved) by overlapping the tiles of half their size and taking only the central half tile, with boundaries at $[x_L/2 : 3x_L/2, y_L/2 : 3y_L/2]$, where $x_L$ and $y_L$ are the horizontal and vertical coordinates of the bottom left corner of a tile. Further improvements can be obtained employing more sophisticated tiling schemes, yet our results show the current solution has only a negligible impact on the results.

The cross-correlation analysis has been adopted as a measure of the effectiveness of the denoiser in ``extracting'' the faintest components of a diffused source, at S/N $<$ 1. The cross-correlation demonstrates how the autoencoder can reliably identify diffused sources in images with strong noise but also complex and bright artefacts, like sidelobes resulting from the aperture synthesis methodology. Extended sources can be tracked to the outermost and faint outskirts by the autoencoder. This is not possible with images processed by the other tested filters, in particular when weak sources are treated. Hence, the autoencoder results to be an effective tool for image segmentation, identifying the pixels belonging to an object as opposed to those of the noisy background. Segmentation is not impacted by the image blurring, and it can guide any further data analysis tool to focus on properly sized regions of interest containing meaningful signals, including areas that would usually be lost being too faint to be identified with other approaches. 

The performance analysis shows that the training time is of the order of hours, depending essentially on the number of epochs and on the training dataset size. The time to process a single image is less than half a second for a 2000$\times$2000 pixels image. The computational time of the autoencoder has an optimal linear scaling with the data size, $\propto N_{\rm image}$, just like the gaussian filter. The FFT algorithm scales slightly worse, $\propto N_{\rm image}$log$(N_{\rm image})$. Therefore the autoencoder can be effectively adopted for increasingly larger images. Furthermore, the implemented tile based approach can lead to the efficient parallelization of the algorithm via domain decomposition on multiple distributed CPUs/GPUs, supporting extremely large images, that cannot fit the memory of a single computing device. 

The adoption of the DDA for denoising and segmentation proves also to have an additional advantage: a properly trained network does not require any further setup or supervision (in particular, human supervision) to process input data. This is a key feature for the processing of huge data volumes. The network can be seamless fed with new incoming data, returning denoised images of regions of space where potential sources lies, together with catalogues of their position, size and possibly other characteristics, like their flux densities. These selected data can then be the target of a focused, in-depth investigation by the astronomer.

The pivotal role of the training data in order to have an effective Machine Learning algorithm emerges clearly from our study. Autoencoders have the advantage of being unsupervised methods, hence they do not require labelling, which is an highly demanding and error prone data pre-preprocessing activity. However, the training set has to be as much {\it complete} as possible, reproducing with the highest possible fidelity the final data the denoiser will be applied to, in order the generalisation to be successful. We have seen how including different kind of noise and artefacts impacts the results of the autoencoder. We have also tested the usage of a network trained on pure noisy data on data coming from the Clean dataset, with results only partially satisfactory, the accuracy decreasing even on those Clean images that are dominated by random noise. Generalisation is however a key factor for the adoption of the autoencoder for observational data, since simulated observations will be required at least to integrate real observations, insufficient, alone, to accomplish a proper training of the network. A focused study of such generalisation is necessary in order to fully understand the potential of autoencoders in radio interferometry. Unquestionably, an outstanding effort has to be devoted to the implementation of methodologies capable of generating realistic mock observations.  

\section{Conclusions}
\label{sec:conclusions}

New opportunities provided by novel Deep Learning approaches, supported by the availability of increasingly capable computing resources, are finding an optimal test case in modern astronomy (among other research domains), given the flurry of complex data that incoming observing facilities are going to deliver. 
In particular, radio interferometry data, resulting from sophisticated image processing approaches, represents an emerging challenge for any data analysis solution, being affected by both random noise and artefacts with completely different statistical and morphological properties. We have studied the usage of a deep convolutional denoising approach to remove noise and artefacts from radio interferometric data, identifying the presence and estimating the brightness of even the faintest diffused sources. An additional challenge is represented by the enormous data volumes expected in the coming decade from observatories like the SKA, which will require software tools capable of an effective exploitation of HPC solutions and full automation, interactive human intervention and supervision becoming impossible. Machine learning solutions appears to ideally address both requirements. 

Our main achievements can be summarised as follows:
\begin{itemize}
    \item a Deep Learning Denoising Autoencoder method has been implemented and its performance extensively studied on radio interferometric images to verify its effectiveness in identifying and extracting diffused sources in low S/N regimes (down to S/N $\ll$ 1). 
    \item The autoencoder has proved to efficiently remove noise and artefacts preserving the properties of the regions of the sources with S/N of the order or lower than 1. 
    \item However, the convolutional, multi-layer architecture of the encoder introduces numerical blurring which impacts in particular the highest peaks of the flux density  distribution which result to be dimmed. 
    \item Therefore the optimal adoption of the autoencoder seems to be for faint (i.e. at S/N $\sim$ 1 or below) sources and/or for performing data segmentation, with the identification of the regions where sources lie, where then the user can focus the analysis with specific, highly accurate (and computationally expensive) source extraction tools. 
    \item The autoencoder can be effectively used on large data, exploiting efficiently state-of-the-art (hybrid, multi-GPU) HPC systems. 
    \item The autoencoder supports full automation: once trained, it can be used seamless on the input data. 
    \item Deep Learning implies the availability of realistic training data, that have been generated by a combination of cosmological numerical simulations, synthetic sky modelling algorithms and aperture synthesis techniques, which resulted in a open data archive of mock radio observations (see the Data Availability section below). The simulated data have taken the LOFAR HBA instrument as a reference for our tests.
\end{itemize}
Further steps need to be accomplished to make the autoencoder operational for actual observations. First and foremost, an in-depth analysis of the generalisation of a network trained on simulated data to observed images. A very preliminary attempt of generalisation has been presented in the paper, highlighting the implied challenges. Such study is already on-going and will be presented in a specific paper (in preparation). Second, a parallel distributed version of the encoder will be developed, in order to support extremely large images like those expected for, e.g., LOFAR VLBI or the SKA (up to $10^5 \times 10^5$ pixels). 

\section*{Data Availability}
\label{data}
The Sky, Clean and Dirty images produced and adopted for this work are publicly available in FITS format at https://owncloud.ia2.inaf.it/index.php/s/IbFPlCCcPUresrr.

\section*{Acknowledgements}
We thank Henrik Edler, Francesco de Gasperin and Nicola Locatelli for the useful discussion and suggestion regarding the generation of the Sky Models and the synthetic observations. We thank also Tim Dykes and Baerbel Koribalski for the feedbacks on the manuscript. 

The cosmological simulations were performed with the {\enzo} code (http://enzo-project.org), which is the product of a collaborative effort of scientists at many universities and national laboratories. We gratefully acknowledge the {\enzo} development group for providing extremely helpful and well-maintained on-line documentation and tutorials.
F.V. acknowledges financial support from the ERC  Starting Grant "MAGCOW", no. 714196.  

The simulations on which this work is based have been produced on Piz Daint supercomputer at CSCS-ETHZ (Lugano, Switzerland) under projects s701 and s805 and on the Marconi100 Supercomputer at CINECA (Bologna, Italy) under project INA17\_C5A38 and INA21\_C8B49 in numerical projects with F.V. as PI. 
We also acknowledge the usage of online storage tools kindly provided by the INAF Astronomical Archive (IA2) initiative (http://www.ia2.inaf.it).  The training and the tests of the autoencoder have been performed on the Marconi100 system of CINECA (Italy) under the grant IscrC\_RadioGPU.

\bibliographystyle{mnras}
\bibliography{franco,franco2}

\appendix
\section{Architecture of the Autoencoder}
\label{sec:appA}

In this Appendix, we investigate how the accuracy of the DDA depends on the architecture of the autoencoder network focusing on two main features: the size of the dense layer and the number of hidden convolutional layers (the depth of the network). The number of features maps is instead kept constant for each test and equal to 32 at level 1 ($128^3$ pixels tiles), 64 at level 2 ($64^3$ pixels tiles), 64 at level 3 ($32^3$ pixels tiles) and 128 at level 4 ($16^3$ pixels tiles). Hyperparameters have been set to $\mu = 0.0001$, $B = 50$, $\epsilon = 100$. We have used the Sky images only, without any additional noise or artefacts, in order to easily highlight the specific impact of the encoder architecture on the data. Accuracy has been estimated using both the mean squared error (MSE, Equation \ref{eq:mse}) and the energy distribution introduced in Section \ref{sec:identity}.

The network architectures reported in Table \ref{tab:a1} have been tested. The presented MSE values are in code units. They show small variations among the different cases.
Nevertheless, accuracy clearly tends to be higher increasing the size of the dense layer. This, in fact, learns from all the combinations of the features of the previous layer and it is the key component to extract relevant information. However, increasing the number of neurons in the dense layer rises also the capacity of the network to learn patterns, making the network lazy and replicating the input values to the output values. This means that, when noisy images are considered, the network tends to learn the noise without extracting any feature: an effect known as {\it overfitting}. Furthermore, the bigger the dense layer, the higher the computational requirements of the encoder. Hence, an optimal trade-off between accuracy and overfitting results to be a dense layer of several hundreds of neurons. 

Considering the flux density distribution, we have calculated the quantity
\begin{equation}
    \Delta_c = \sum^{bins}_{i=1} \vert E_{{\rm sky},i} - E_{{\rm auto},i}\vert,
\end{equation}
where $E_{{\rm sky},i}$ and $E_{{\rm auto},i}$ are the number of counts in the i-th flux density bin for the Sky models and the encoded images respectively. The quantity $\Delta_c$ is the {\it discrepancy} between the true and reconstructed energy distributions. The highest its value, the worse is the quality of the encoding. 

\begin{table}
\centering
\begin{tabular}{c|c|c|c|c}
Hidden Layers & Dense Layer & MSE$_{3\times3}$ & $\Delta_c$ &MSE$_{5\times5}$\\\hline\hline
2 & 50 & 3.246 & 191610 & 3.218\\
2 & 100 & 1.229 & 49068 & 1.263\\
2 & 200 & 0.575 & 59180 & 0.598\\
2 & 400 & 0.400 & 62300 & 0.435\\
3 & 50 & 3.259 & 141288 & 3.245\\
3 & 100 & 1.209 & 312606 &  1.111\\
3 & 200 & 0.605 & 263144 & 0.543\\
3 & 400 & 0.611 & 199036 & 0.670\\
4 & 50 & 5.017 & 86408 & 2.677\\
4 & 100 & 1.407 & 241464 & 2.008\\
4 & 200 & 0.934 & 130892 & 0.876\\
4 & 400 & 0.854 & 63698 & 2.313\\
\hline
\end{tabular}
\caption{The first and the second columns represent the depth of the network and the size of the dense layer respectively. The third and the fifth columns show the mean squared error for the different network configurations, with a receptive window of $3\times3$ and $5\times5$ pixels respectively. The MSE is in code units. The fourth column shows the discrepancy parameter. Convolutional hidden layers accounts for 32, 64, 64, 128 features maps, from those at the highest to those at the lowest resolution.}
\label{tab:a1}
\end{table}

\begin{figure*}
\begin{center}
\includegraphics[width=0.95\textwidth]{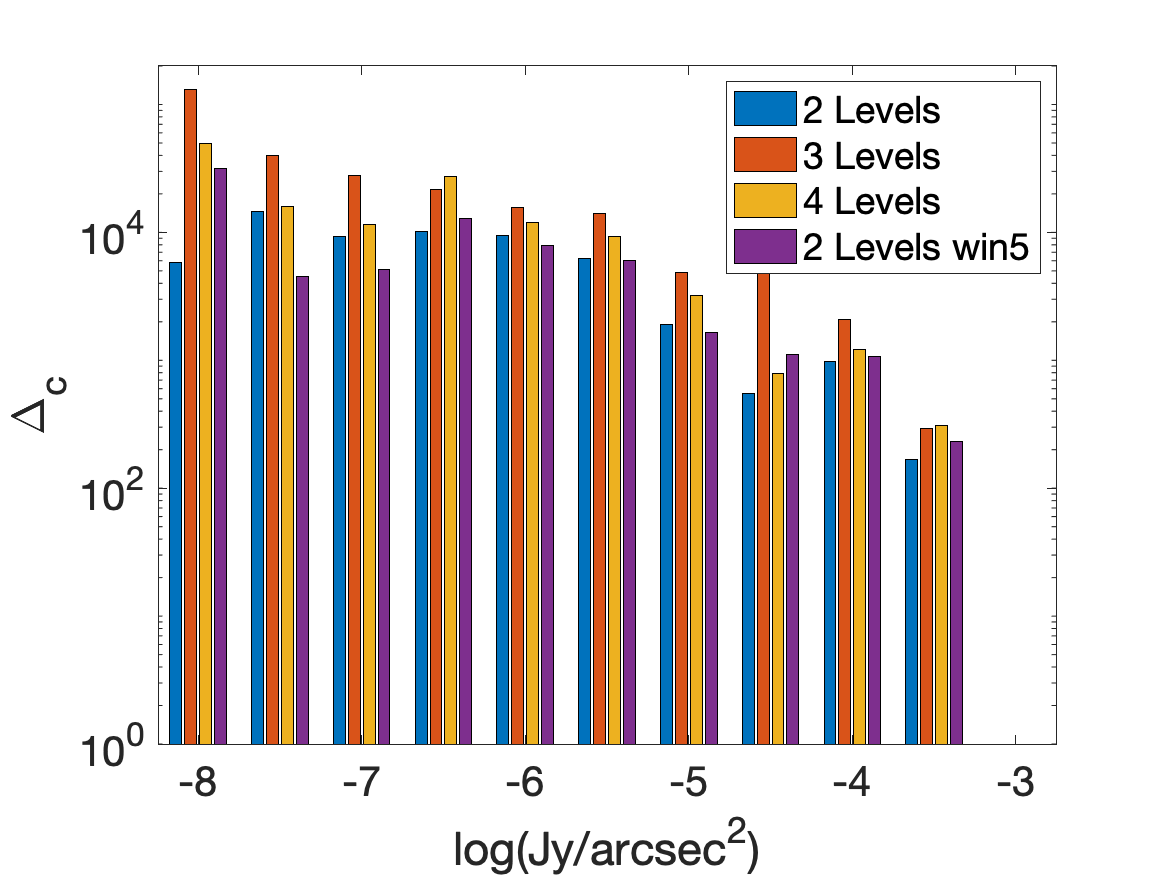}
\caption{Discrepancy distribution for three different architectures with 2, 3 and 4 hidden layers respectively (first three bars of each energy bin). The win5 test is the same as the 2 Levels test but with a convolutional window of 5$\times$5 pixels (fourth bar).}
\label{fig:histodiff}
\end{center}
\end{figure*}

The combination of the MSE and of the discrepancy shows how the 2 levels networks gives better results compared to the others. In Figure \ref{fig:histodiff} we focus on such case presenting the discrepancy distribution for the networks with  different depths (first three bars of each energy bin). In almost all cases, the 2 levels network gives the best results, followed by the 4 levels one. Three levels produce in this case the worst results. We have then repeated the tests increasing the size of the receptive window (the kernel used for convolution) from $3\times3$ to $5\times5$ pixels (``win5'' cases). The resulting values of MSE are, in most cases, compatible with those obtained using the smallest window. However, looking at the discrepancy distribution in the 2 levels, 200 dense neurons case, it is evident how, at flux densities $>10^{-5}$Jy/arcsec$^2$, the win5 kernel gives worse results than the $3\times3$ pixels ones. This corresponds to the drop of brightness observed in the highest peaks of the flux density maps presented in the Section \ref{sec:results}. It can be interpreted as caused by numerical diffusion introduced by the convolution operation at each hidden layer. 

Summing up, our tests suggest an architecture consisting in two convolutional plus two pooling layers for encoding, two convolutional with two unpooling layers for decoding and one dense layer made of 200/400 neurons as the most effective for our data. The 200 elements dense layer is to prefer for noisy data to avoid overfitting. The loss of brightness in the highest peaks of the flux density distribution can be ascribed to an unavoidable numerical blurring effect of the convolutional approach. 

\section{Hyperparameters Tuning}
\label{sec:appB}

A number of tests have been performed in order to identify the ideal hyperparameters set-up for the DDA, adopting the architecture identified in Appendix A as the most effective. Due to the stochastic nature of the algorithm, it is impossible to select a combination of hyperparameters producing a single best result. Our analysis aims to determine for each hyperparameter, the range of values producing the most accurate results. In all the tests that follow, the autoencoder has been trained for all the combinations of the following hyperparameters values:
\begin{itemize}
    \item[] $\mu = [0.1, 0.01, 0.001, 0.0001, 0.00005, 0.00001]$;
    \item[] $B = [25, 50, 100, 150, 200]$;
    \item[] $\epsilon = [50, 100, 300, 500].$
\end{itemize}
As a measure of the accuracy we use the MSE. In addition, in order to select the best setup in the Identity case (Section \ref{sec:identity}) we have also introduced the mean relative error, that allows to have a meaningful estimate of the accuracy of the result in the case of images dominated by $\sim$0 flux pixels. 

For the Identity test we obtain the values reported in Table \ref{tab:identitytest}, which is limited to the best 10 results. Among these results there are no hits with learning rate $\mu = 0.01$ and $\mu = 0.00001$, and the the optimal range of values for this parameter is between 0.00005 and 0.001, with no particular preference for one of a specific value, although the minimum MSE is obtained for $\mu = 0.0001$. A further conclusion is that a number of epochs equal to 50 is not sufficient for the training to converge. Number of epochs of 100 and 300 give comparable results. The former value is anyway preferable since it leads to a converged network in a shorter time. On the other hand, a large number of epochs could lead to overfitting (in particular when noisy data will be treated) and has to be avoided. The optimal batch size is obtained with values of 50 and 100. Only in one case a smaller batch size, $B=25$, is present in the top 10 list, while $B\ge 150$ cases never appear. In order to select the final setup for the Identity case, the MSE has been complemented in Section \ref{sec:identity} with the mean relative error (Equation \ref{eq:mre}), that allows to have a meaningful estimate of the accuracy of the result in the case of images dominated by $\sim$0 flux pixels. 

\begin{table}
\centering
\begin{tabular}{c|c|c|c}
$\epsilon$ & B & $\mu$ & MSE \\\hline\hline
100&	50&	    0.0001&	0.382\\	
300&	50&	    0.00005&	0.412\\		
100&	25&	    0.0001&	0.415\\		
300&	100&	0.005&	0.432\\		
300&	100&	0.0001&	0.438\\		
100&	50&	    0.005&	0.449\\		
100&	100&	0.001&	0.479\\		
300&	100&	0.001&	0.489\\		
100&	50&	    0.00005&	0.535\\		
100&	50&	    0.001&	0.548\\		
\hline
\end{tabular}
\caption{Mean squared error in code units, for the Identity test as a result of the training of the autoencoder using the values of the hyperparameters reported in the first three columns (best 10 hits are presented).}
\label{tab:identitytest}
\end{table}

Table \ref{tab:randomtest} show the top 10 MSEs for the case of images with Random Noise. Also in this case, 50 epochs are clearly not enough to reach convergence. On the other hand, as expected, 300 epochs tend to overfit the data, hence they lead, in general, to a loss of accuracy. Hence, the value $\epsilon = 100$ appears to be the most effective for this dataset. A tendency toward batch sizes bigger than for the Identity case, between 100 and 150 ($B=200$, instead, being less effective for the training) emerges. However, visual inspection and the analysis of the flux density distribution (not presented here), show how larger batch sizes tend to be more diffusive than smaller one, leading to bigger drops of the brightness of the highest peaks of the distribution. Therefore, we have kept a batch size of 50.
Regarding learning rate, values bigger than 0.001 and smaller than 0.00005 are ruled out, while within this range the algorithm typically converge with similar MSE values. 

\begin{table}
\centering
\begin{tabular}{c|c|c|c}
$\epsilon$ & B & $\mu$ & MSE \\\hline\hline
100&	100&	0.00100&	27.181\\
100&	150&	0.00100&	28.348\\
100&	150&	0.00005&	29.531\\
100&	50&	    0.00010&	29.546\\
100&	50&	    0.00005&	29.784\\
100&	100&	0.00010&	30.007\\
100&	200&	0.00010&	30.525\\
100&	100&	0.00005&	30.526\\
300&	100&	0.00010&	30.738\\
100&	100&	0.00010&	30.769\\
\hline
\end{tabular}
\caption{Mean squared error in code units for the Random Noise test, as a result of the training of the autoencoder using the values of the hyperparameters reported in the first three columns (best 10 hits are presented).}
\label{tab:randomtest}
\end{table}

For Clean images, the best results are presented in Table \ref{tab:cleantest}. The main difference with the previous cases is that a much larger number of epochs, $\epsilon$ = 500 is required by the algorithm to converge. The optimal batch size and learning rate are instead analogous to the previous case. Similar results (not presented in details) are found for the Dirty dataset. 

\begin{table}
\centering
\begin{tabular}{c|c|c|c}
$\epsilon$ & B & $\mu$ & MSE \\\hline\hline
500&     50&      0.0001&  45.882\\
500&     100&     0.001&   49.694\\
300&     50 &     0.001&   52.817\\
500&    50  &    0.00005& 62.978\\
300&     100&     0.001&   87.378\\
300&     50 &     0.0001&  90.731\\
500&     100&     0.0001&  106.497\\
100&     100&     0.00005& 136.113\\
300&     50 &     0.00005& 155.270\\
300&     100&     0.0001&  178.352\\
\hline
\end{tabular}
\caption{Mean squared error in code units for the Clean test, as a result of the training of the autoencoder using the values of the hyperparameters reported in the first three columns (best 10 hits are presented).}
\label{tab:cleantest}
\end{table}
\end{document}